\documentclass{elsart}

\usepackage[square,comma]{natbib}
\usepackage{graphicx}
\usepackage{pxfonts}
\usepackage{lineno}

\usepackage{amssymb}

\journal{}

\begin{document}

\thispagestyle{empty}
\begin{Large}
\textbf{DEUTSCHES ELEKTRONEN-SYNCHROTRON}

\textbf{\large{Ein Forschungszentrum der
Helmholtz-Gemeinschaft}\\}
\end{Large}

DESY 10-080

June 2010

\begin{eqnarray}
\nonumber &&\cr \nonumber && \cr \nonumber &&\cr
\end{eqnarray}
\begin{eqnarray}
\nonumber
\end{eqnarray}
\begin{center}
\begin{Large}
\textbf{Cascade self-seeding scheme with wake monochromator for narrow-bandwidth X-ray FELs}
\end{Large}
\begin{eqnarray}
\nonumber &&\cr \nonumber && \cr
\end{eqnarray}

\begin{large}
Gianluca Geloni,
\end{large}
\textsl{\\European XFEL GmbH, Hamburg}
\begin{large}

Vitali Kocharyan and Evgeni Saldin
\end{large}
\textsl{\\Deutsches Elektronen-Synchrotron DESY, Hamburg}
\begin{eqnarray}
\nonumber
\end{eqnarray}
\begin{eqnarray}
\nonumber
\end{eqnarray}
ISSN 0418-9833
\begin{eqnarray}
\nonumber
\end{eqnarray}
\begin{large}
\textbf{NOTKESTRASSE 85 - 22607 HAMBURG}
\end{large}
\end{center}
\clearpage
\newpage

\begin{frontmatter}



\title{Cascade self-seeding scheme with wake monochromator for narrow-bandwidth X-ray FELs}


\author[XFEL]{Gianluca Geloni\thanksref{corr},}
\thanks[corr]{Corresponding Author. E-mail address: gianluca.geloni@xfel.eu}
\author[DESY]{Vitali Kocharyan}
\author[DESY]{and Evgeni Saldin}

\address[XFEL]{European XFEL GmbH, Hamburg, Germany}
\address[DESY]{Deutsches Elektronen-Synchrotron (DESY), Hamburg,
Germany}

\begin{abstract}
Three different approaches have been proposed so far for
production of highly monochromatic X-rays from a baseline XFEL
undulator: (i) single-bunch self-seeding scheme with a four crystal
monochromator in Bragg reflection geometry; (ii) double-bunch
self-seeding scheme with a four-crystal monochromator in Bragg
reflection geometry; (iii) single-bunch self-seeding scheme with
a wake monochromator. A unique element of the X-ray optical design
of the last scheme is the monochromatization of X-rays using a
single crystal in Bragg-transmission geometry. A great advantage
of this method is that the monochromator introduces no path delay
of X-rays. This fact eliminates the need for a long electron beam
bypass, or for the creation of two precisely separated, identical
electron bunches, as required in the other two self-seeding
schemes. In its simplest configuration, the self-seeded XFEL
consists of an input undulator and an output undulator separated
by a monochromator. In some experimental situations this simplest
two-undulator configuration is not optimal. The obvious and
technically possible extension is to use a setup with three or
more undulators separated by monochromators. This
amplification-monochromatization cascade scheme is distinguished,
in performance, by a small heat-loading of crystals and a high spectral purity of the output radiation. This
paper describes such cascade self-seeding scheme with wake
monochromators. We present feasibility study and exemplifications
for the SASE2 line of the European XFEL.
\end{abstract}

%
%
%
\end{frontmatter}



\section{\label{sec:intro} Introduction}

The self-seeding approach in Free-Electron Lasers (FELs) was
proposed to obtain a bandwidth narrower than that achievable with
conventional X-ray SASE schemes \cite{tdr-2006}-\cite{SPRIN}. A
self-seeded FEL consists of two undulators with a monochromanor
located between them. In the original VUV-soft X-ray
case, a grating monochromator was proposed \cite{SELF}. Three types of self-seeding schemes for hard
X-ray FELs have been studied \cite{SXFE}-\cite{OURX}.
Historically, the first type was a single bunch self seeding
scheme with a four-crystal monochromator in Bragg reflection
geometry \cite{SXFE}. A second type was the double bunch
self-seeding scheme with four crystal monochromator in Bragg
reflection geometry \cite{OURL,HUAN}, and a third type was a
single-bunch self-seeding scheme making use of a wake
monochromator \cite{OURX}. A unique element of the X-ray optical
design of the last scheme is the monochromatization of X-rays
using a single crystal in Bragg transmission geometry. The X-ray
beam is transmitted through a thick crystal oriented for Bragg
reflection. A great advantage of this method is that it introduces
no path-delay of X-rays in the monochromator, thus avoiding the
need for a long electron beam bypass, or for the creation of two
precisely separated, identical electron bunches, as
required in schemes with the monochromator in Bragg reflection
geometry.

In \cite{OURX} we discussed how such kind of self-seeding scheme
may be combined with a fresh bunch technique
\cite{HUAYU}-\cite{OUR03}. The combination of self-seeding and
fresh bunch techniques is extremely insensitive to noise and
non-ideal effects. In fact, the radiation pulse used to produce
the monochromatic (wake) seed signal is in the GW-level power.
This large power can tremendously improve the signal-to-noise
ratio of the self-seeding scheme. The possibility of combining
self-seeding scheme and fresh bunch technique would be of great importance especially during the early experimental stage, when a proof of principle is built.

Despite of these advantages, the combination of self-seeding and
fresh bunch techniques suffers from two drawbacks. First, there is
a strong SASE signal outside of the seed bandwidth, due to
application of fresh bunch technique\footnote{Actually this can be
transformed into an advantage. The SASE pulse can be separated
from the longitudinally coherent seed pulse by the post
monochromator in the experimental hall and exploited separately.}. Second,
the power of the SASE pulse which impinges on the crystal is
relatively large and heat-loading problems are not automatically
avoided. This is the case, in particular, when the combination of
self-seeding and fresh bunch techniques is used at the European
XFEL, which is characterized by a very high repetition rate. These
drawbacks can be overcome by using the cascade self-seeding scheme
proposed in the present work. In this paper we study such scheme,
which consists of two parts: a succession of two
amplification-monochromatization cascades and an output undulator
in which the monochromatic seed signal is amplified up to saturation.
Each cascade consists of an undulator, acting as an amplifier, and
a monochromator.

In the next section we describe in detail the method and the basic
principles of our technique. Subsequently, in the following
sections, we study the feasibility of the technique, giving
exemplification for the SASE2 beamline at the European XFEL.

\section{Principle of cascade self-seeding techniques based on the use of wake monochromators}

\begin{figure}[tb]
\includegraphics[width=1.0\textwidth]{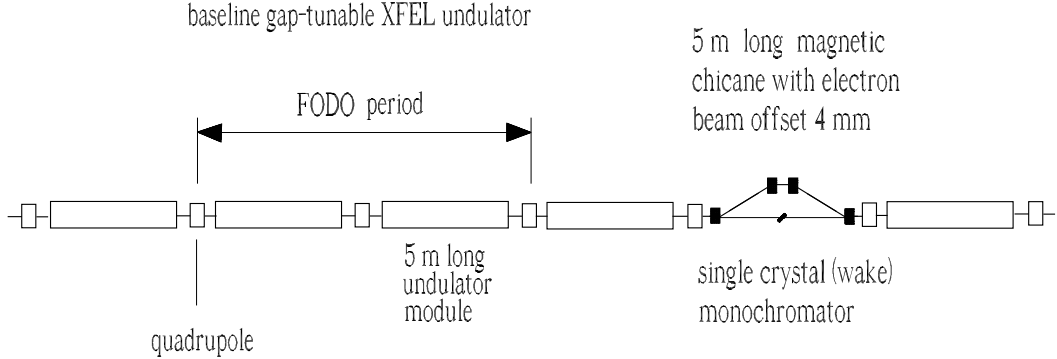}
\caption{Installation of a wake monochromator in the baseline XFEL
undulator. The setup is composed of two components, a crystal and
a weak magnetic chicane. The magnetic chicane accomplishes three
tasks by itself. It creates an offset for crystal installation, it
removes the electron micro-bunching produced in the upstream
undulator, and it acts as a delay line for temporal windowing. The
transmittance spectrum of the crystal shows a narrow-band
absorption resonance when the incident X-ray beam satisfies the
Bragg diffraction condition. The temporal waveform of the
transmitted radiation pulse is characterized by a long
monochromatic wake. After the crystal, the monochromatic wake of
the radiation pulse is combined with the delayed electron bunch,
and amplified in the downstream undulator.} \label{cs1}
\end{figure}
\begin{figure}[tb]
\includegraphics[width=1.0\textwidth]{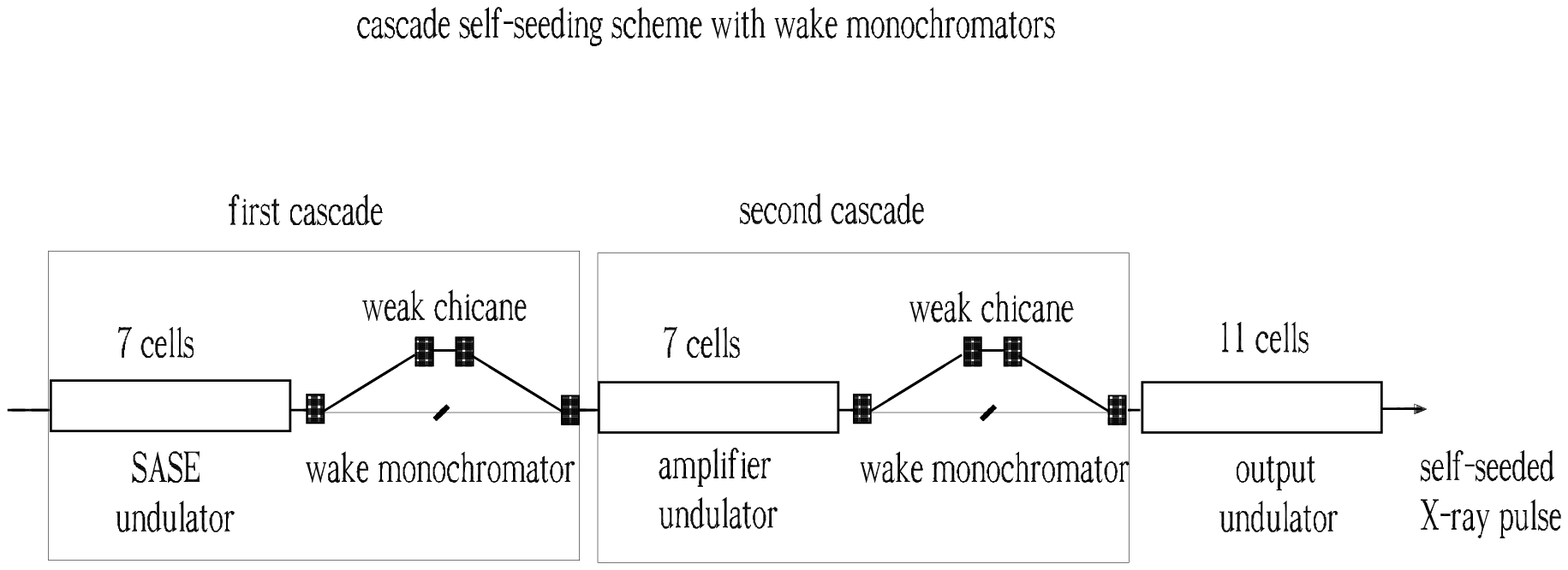}
\caption{Design of an undulator system for narrow-bandwidth mode
of operation. The scheme is based on the use of a cascade, single
bunch self-seeding scheme with wake monochromators. In
performance, the cascade type of self-seeded XFEL  is
distinguished by its small heat-loading of monochromators.}
\label{cs2}
\end{figure}
\begin{figure}[tb]
\includegraphics[width=1.0\textwidth]{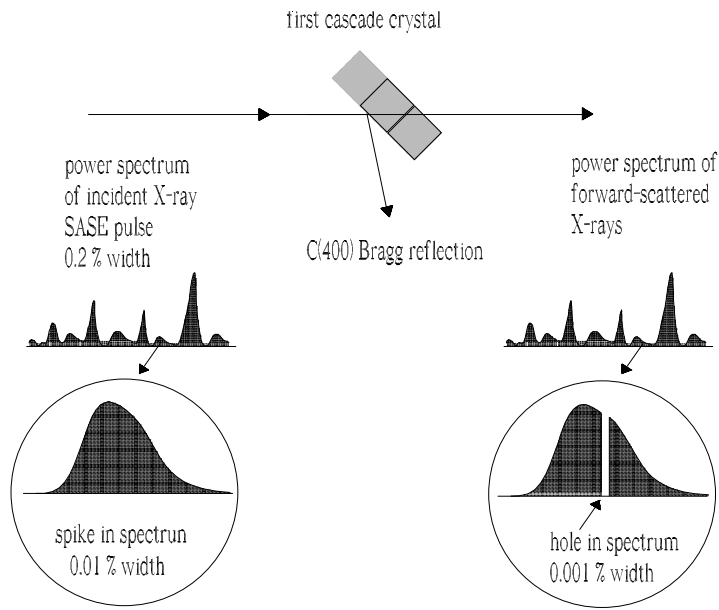}
\caption{First cascade. Short pulse mode of operation. The crystal
acts as a bandstop filter for the transmitted X-ray SASE radiation
pulse.} \label{cs3}
\end{figure}
The self-seeding technique considered in this work is based on the
substitution of a single undulator module with a weak chicane and
a single crystal, as shown in Fig. \ref{cs1}. Two cascades can be
arranged sequentially as shown in Fig. \ref{cs2}. In this paper we
will consider both the case when a single cascade or two cascades
are present. With reference to Fig. \ref{cs2}, we begin to study
the case when two-cascades are present.

The first undulator in Fig. \ref{cs2} operates in the linear
high-gain regime starting from the shot-noise in the electron
beam. After the first undulator, the output SASE radiation passes
through the monochromator, which reduces the bandwidth to the
desired value. According to the wake monochromator principle, the
SASE pulse coming from the first undulator impinges on a crystal
set for Bragg diffraction. Then, the single crystal in Bragg
geometry actually operates as a bandstop filter for the
transmitted X-ray SASE radiation pulse, as shown in Fig.
\ref{cs3}. When the incident angle and the spectral contents of
the incoming beam satisfy the Bragg diffraction condition, the
temporal waveform of the transmitted radiation pulse shows a long
monochromatic wake. The duration of this wake is inversely
proportional to the bandwidth of the absorption line in the
transmittance spectrum.

While the radiation is sent through the crystal, the electron beam
passes through a magnetic chicane, which accomplishes three tasks
by itself: it creates an offset for the crystal installation, it
removes the electron micro-bunching produced in the first
undulator, and it acts as a delay line for the implementation of
the temporal windowing. In other words, the magnetic chicane
shifts the electron bunch on top of the monochromatic wake created
by the bandstop filter thus selecting (temporal windowing) a part
of the wake. By this, the electron bunch is seeded with a
radiation pulse characterized by a bandwidth much narrower than
the natural FEL bandwidth.

For the hard X-ray wavelength range, a small dispersive strength  $R_{56}$  in the order of ten microns is sufficient to remove the micro bunching in the electron bunch.  As a result, the choice of
the strength of the magnetic chicane only depends on the delay that we want to introduce between electron bunch and radiation. In our case, this amounts to $6 ~\mu$m for the short pulse mode of operation and to $60~\mu$m  for the long pulse mode of operation. Such dispersion strength is small enough to be generated by a short $5$ m-long chicane to be installed in place of a single undulator module. Such chicane is, however, strong enough to create a sufficiently large transverse offset of a few millimeters for installing the crystal.

Successful operation of the self-seeded XFEL requires fulfillment
of several requirements. The first undulator must operate in the
deep linear regime, and not in saturation. In fact, the
amplification process in the FEL leads to an energy modulation in
the electron beam. After the electron beam passes through the
magnetic chicane, such energy modulation transforms into
additional energy spread. Calculations show that in order not to
spoil the electron beam quality, the power gain of the first
undulator should be about three orders of magnitude smaller than
the power gain of the X-ray SASE FEL at saturation. Here we
consider the hard X-ray mode of operation, where the effective
power of shot-noise in the electron beam is $P_\mathrm{n} \sim 3$
kW, and the power of the SASE radiation at saturation is
$P_\mathrm{sat} \sim 30$ GW. It follows that for effective
operation of the second undulator, one requires that the power
gain of the first undulator be no more than four orders of
magnitude.

To provide effective operation of the self-seeded XFEL, we also
require that the power of the monochromatic seed signal
$P_\mathrm{seed}$ at the entrance of the output undulator
significantly exceeds the effective shot-noise power in the
electron beam, i.e. $P_\mathrm{seed} \gg P_\mathrm{n}$. Calling
$G_1$ the power gain in the first undulator, and $T_m$ the
transmission factor of the monochromator one has
$P_\mathrm{seed1}/P_{n} \sim G_1 T_m$. For a bandpass filer the
transmission factor is simply the ratio of the of the transmitted
pulse power to the incoming pulse power. Similarly, for the wake
monochromator we can define $T_m$ as the ratio of the wake power
in the temporal window defined by the electron bunch length to the
incoming pulse power. More in detail, $T_m$ can be written the
product of a geometrical factor $R_m$ and a coefficient $K_s$,
i.e. $T_m \sim R_m K_s$. The geometrical factor $R_m$ depends on
the line width of the bandstop filter, $\Delta \omega_m$, and on
the time delay $\tau_d$, while the coefficient $K_s$ can be
written in the form $K_s \sim \Delta \omega_m/(\Delta
\omega_{in})$, where $\Delta \omega_{in}$ is the bandwidth of the
incoming X-ray beam. Using only a single cascade, i.e. considering
the simplest possible two-undulator configuration of the
self-seeded XFEL, the bandwidth of the incoming beam is actually
the radiation bandwidth of the SASE XFEL i.e. $\Delta \omega_{in}
\sim \Delta \omega_{SASE}$. Our calculations (see the next Sections) show that in this case the transmission factor for the wake monochromator is about $0.1
\%$. Remembering that $P_\mathrm{seed1}/P_{n} \sim G_1 T_m$, and that
$G_1 \lesssim 10^{4}$, one cannot fulfill the requirement
$P_\mathrm{seed1} \gg P_\mathrm{n}$. In \cite{OURX} we proposed a
method to get around this obstacle. It is based on the application
of the self-seeding scheme with wake monochromator described above
in combination with a fresh bunch technique. At variance, here we
propose a new method to increase the value of $P_\mathrm{seed}$. A
high value of the signal-to-noise ratio at the entrance of the
output undulator may in fact be obtained by using more than one
stage of amplification-monochromatization.

\begin{figure}[tb]
\includegraphics[width=1.0\textwidth]{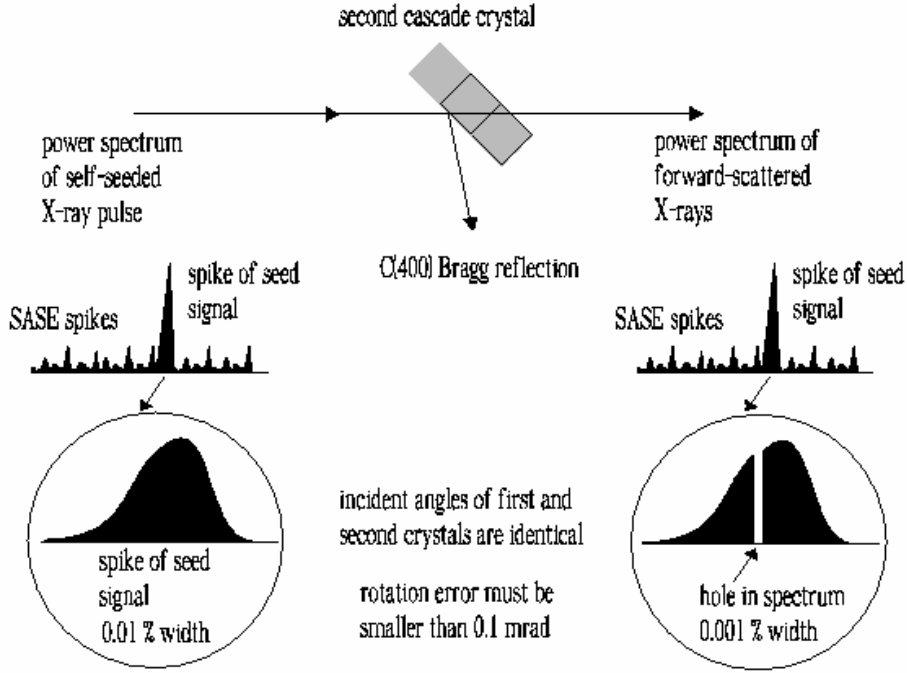}
\caption{Second cascade. The crystal acts as a bandstop filter for
the self-seeded X-ray pulse. The incident angles of first and
second crystals are identical. The rotation error must be smaller
than $0.1$ mrad for the short pulse mode of operation.}
\label{cs4}
\end{figure}
\begin{figure}[tb]
\includegraphics[width=1.0\textwidth]{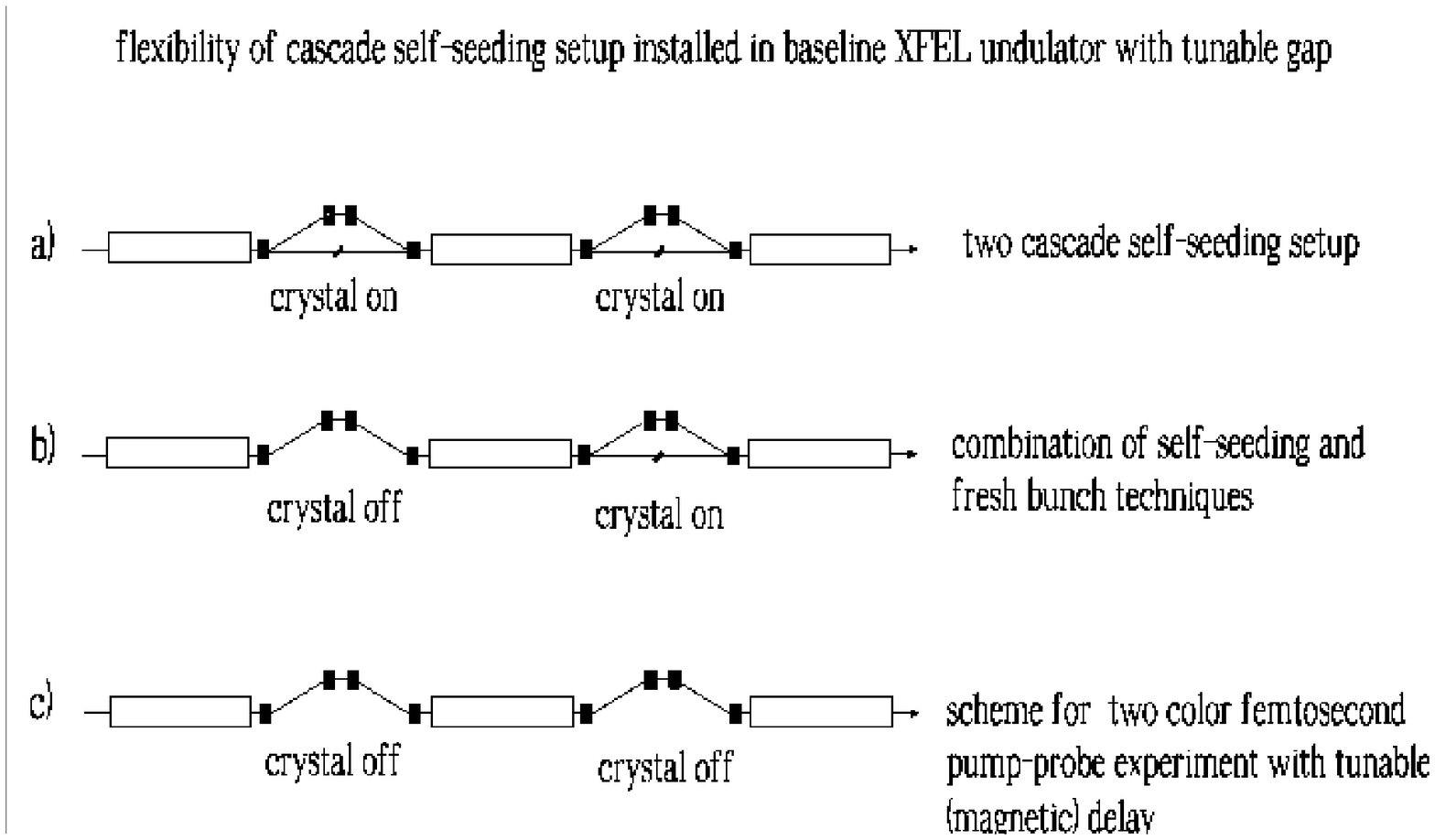}
\caption{Three modes of operation for the two weak magnetic
chicanes installed in the XFEL baseline undulator with tunable gap
are foreseen.} \label{cs5}
\end{figure}
Consider the three-undulators self-seeding scheme (i.e. two
cascades) as shown in Fig. \ref{cs2}. To be specific, here we
consider the case of two identical cascades. As discussed above,
the signal-to-noise ratio at the entrance of the second undulator
cannot be much larger than unity. Nevertheless,
the monochromatic field amplitude at the entrance of the third
undulator will be much larger than that at the entrance of the second. The difference with respect to the previously discussed two-undulator configuration is that after amplification in the second undulator, Fig. \ref{cs4}, the bandwidth of the X-ray pulse
related with the monochromatic signal that impinges on the second
crystal is near to the transform-limited bandwidth of the electron
bunch i.e. $c/\sigma_e$. In other words, the transmission factor is now $T_m = R_m \Delta \omega_m c/\sigma_e$.
As a result, assuming the same amplification $G_2 = G_1$, the signal to noise ratio is enhanced by a factor
given by the ratio of the two transmission factors, i.e. $\sigma_e \Delta \omega_{SASE}/c$. A rough
estimate for the signal to noise ratio at the entrance of the
third (output) undulator is therefore $P_\mathrm{seed2}/P_\mathrm{n} \sim (P_\mathrm{seed1}/P_\mathrm{n})
\sigma_e \Delta \omega_{SASE}/c \gg 1$. Since this value is much larger than unity, we
conclude that a double cascade self-seeding scheme using a
wake monochromator is insensitive to noise and non-ideal effects.

It should be noted that the three-undulator configuration in Fig.
\ref{cs2} can be naturally taken advantage of in different
schemes, as shown in Fig. \ref{cs5}. The upper figure (a) shows
the present self-seeding scheme. The middle figure (b) refers to
the self-seeding scheme in combination with the fresh-bunch
technique, as discussed in \cite{OURX}. Finally, the lower figure
(c) shows the advantage of a two-chicane setup when dealing with
pump-probe techniques. In this case, as considered in
\cite{OUR01}, the first chicane enforces a fresh bunch technique
and prepares a radiation pulse of a given color while, the second
can be used to delay the electron bunch relative to such pulse, to
obtain a tunable delay. The three above-mentioned setups are now
combined in a single unit composed by three undulators
equipped with two wake monochromators.

\begin{figure}[tb]
\includegraphics[width=1.0\textwidth]{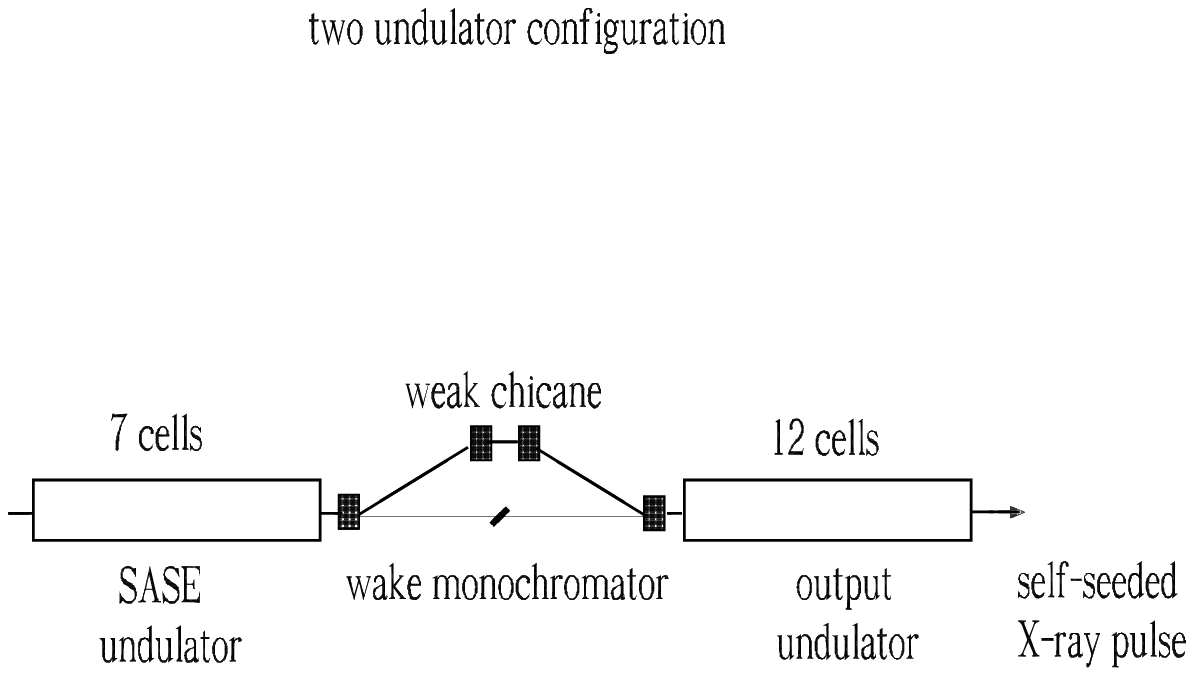}
\caption{Self-seeding setup with wake monochromator. Two undulator
configuration. Short pulse mode of operation. Compared with the
two cascade case, where almost all photons radiated within the
seed bandwidth, here there is strong output  SASE radiation
outside this bandwidth. }
\label{cs6}
\end{figure}

Finally, it is interesting to briefly discuss the simpler two-undulator configuration, Fig. \ref{cs6}. The two-undulator configuration may be particularly advantageous in cases when the total available undulator length is too short to enforce the three-undulator configuration. In this case, one may consider an increase in length of the first undulator in order to increase the contrast between seeded and SASE signal, at the cost of increasing the heat load.

As discussed above, in the two-undulator configuration,  even if the instantaneous powers of seed signal and  SASE noise are comparable, in the Fourier domain there is an enhance in photon spectral density order of $\sigma_e \Delta \omega_{SASE}/c >> 1$, because of the different spectral widths of seed and SASE noise. As a result, even though we have no spectral purity, in the case of two-undulator configuration the spectral density  is still comparable with the three-undulator configuration. If it is possible to use a post-monochromator to filter out the SASE signal, a user will obtain a radiation source with characteristics comparable to the three-undulator configuration, with larger fluctuations and a few times smaller brightness.  Spectral purity can be, however, of crucial importance for application of some other
techniques like tapering and polarization manipulations, i.e. for the next steps of performance improvement.

\section{Heat Loading issues}

In a previous paper of us \cite{OURX} we proposed to couple the monochromatization through a wake monochromator with a fresh-bunch technique. As noted above, this solution has advantages in terms of high signal to noise ratio, but also disadvantages, among which the presence of off-band SASE radiation, and a high heat-load of the monochromator crystal. For the European XFEL, the combination of our self-seeding scheme with a fresh bunch technique, as proposed in \cite{OURL}, satisfies the heat loading restrictions for the average power density. In fact, we showed that the situation is not different compared to third generation sources. However, the European XFEL differs compared to third generation sources in the very specific time diagram, which foresees the production of about $10$ trains of pulses per second, each train consisting of about $3000$ pulses. In this case, the average power density along a single pulse-train is the meaningful figure of merit, rather than the above-mentioned average power density.

The two-cascade setup drastically relaxes the heat load on the
crystals, and eliminates the presence of the off-band SASE radiation, still retaining the advantage of a high signal to noise ratio.
The energy per bunch impinging on the second crystal, which bears
the largest heat-load, can be estimated as $150$ nJ for the short
bunch ($1 \mu$m) mode of operation and $1.5 \mu$J for the long
bunch ($10 \mu$m) mode of operation (see Section \ref{Feas}, Fig.
\ref{IIPwin} and Section \ref{Feal}, Fig. \ref{IIPwin_L}). Let us consider the long-bunch mode of operation, which poses the most difficult challenge. From the previous numbers, one can easily estimate an average power of 50 mW ($1.5 \mu$J $\times 3000$ pulses/train $\times 10$ trains/s). This corresponds to a normal incident power density\footnote{We consider a transverse rms dimension of the bunch of about $20 \mu$m. Assuming, with some approximation, that radiation is distributed as the electron bunch, we obtain an area of $2.4 \cdot 10^{-3} \mathrm{mm^2}$.} of $20 \mathrm{W/mm^2}$ at the position of the second monochromator, already an order of magnitude smaller compared with the average power density at monochromators of third generation synchrotron sources.

The average power within a single bunch train can be estimated by
multiplying the energy by about $3\cdot 10^3$ pulses composing a
single train and dividing by the temporal duration of a train,
which is $0.6$ ms. One obtains a maximum power (for the long bunch
mode of operation) of $7.5$ W. Considering a transverse rms bunch
dimension of about $20 \mu$m, as before, we obtain a power density of about $3$ kW/mm$^2$ within a single train, at normal incidence. Such heat-load is orders of magnitude smaller
than what is foreseen at monochromators for the baseline SASE2, where a diamond crystal with the same thickness ($0.1$ mm) is planned to be used.

\section{\label{Feas} Feasibility study for a short bunch}

Following the previous introduction of the proposed methods we report on a feasibility study of the single-bunch self-seeding scheme. This feasibility study is performed with the
help of the FEL code GENESIS 1.3 \cite{GENE} running on a parallel
machine. In this section we will present the feasibility
study for the short-pulse mode of operation ($1~\mu$m rms) while, later
on, we will cover the long-pulse mode of operation ($10~\mu$m rms). We will treat both the two-undulator as well as the three-undulator configuration. Parameters used in the simulations for the short pulse mode of
operation are presented in Table \ref{tt1}. For the long pulse
mode of operations Table \ref{tt1} is still valid, except for a
ten times larger charge ($0.25$ nC) and a ten times longer rms
bunch length.  We present a statistical analysis consisting of $100$ runs for both short and long pulse mode of operation.

\begin{table}
\caption{Parameters for the short pulse mode of operation used in
this paper.}

\begin{small}\begin{tabular}{ l c c}
\hline & ~ Units &  ~ \\ \hline
Undulator period      & mm                  & 48     \\
K parameter (rms)     & -                   & 2.516  \\
Wavelength            & nm                  & 0.15   \\
Energy                & GeV                 & 17.5   \\
Charge                & nC                  & 0.025 \\
Bunch length (rms)    & $\mu$m              & 1.0    \\
Normalized emittance  & mm~mrad             & 0.4    \\
Energy spread         & MeV                 & 1.5    \\
\hline
\end{tabular}\end{small}
\label{tt1}
\end{table}

\newpage
\subsection{Two-undulator configuration}

\begin{figure}[tb]
\includegraphics[width=1.0\textwidth]{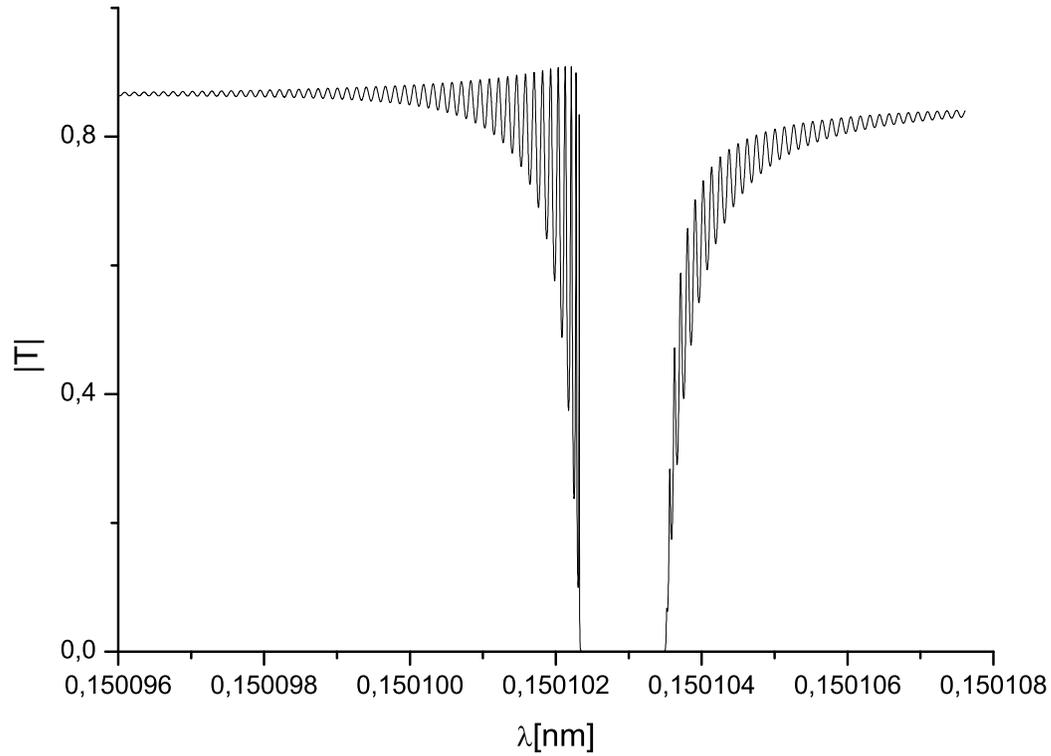}
\caption{Transmissivity (sigma polarization) relevant to the Bragg 400 diffraction of X-rays at $0.15$ nm from a Diamond crystal with a thickness of $0.1$ mm.} \label{ImodT}
\end{figure}

\begin{figure}[tb]
\includegraphics[width=1.0\textwidth]{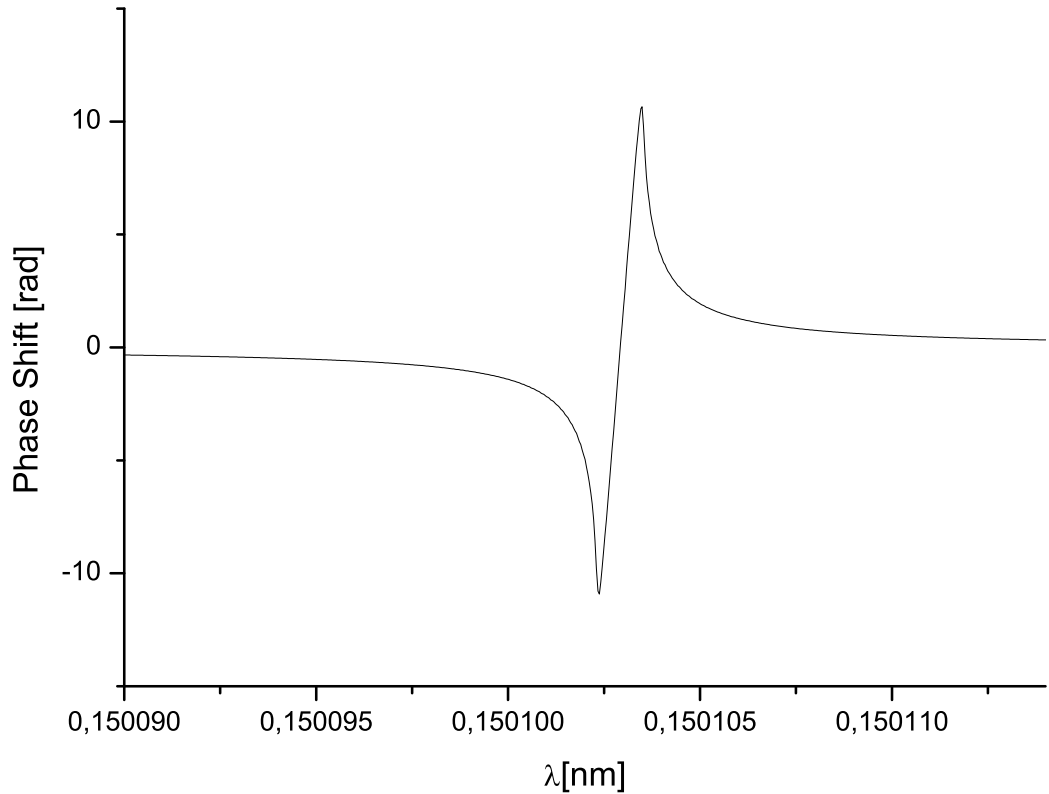}
\caption{Minimal phase shift of the forward-diffracted X-rays at
$0.15$ nm (sigma polarization) relevant to the Bragg 400
diffraction from a Diamond crystal with a thickness of $0.1$ mm.} \label{IphT}
\end{figure}

\begin{figure}[tb]
\includegraphics[width=1.0\textwidth]{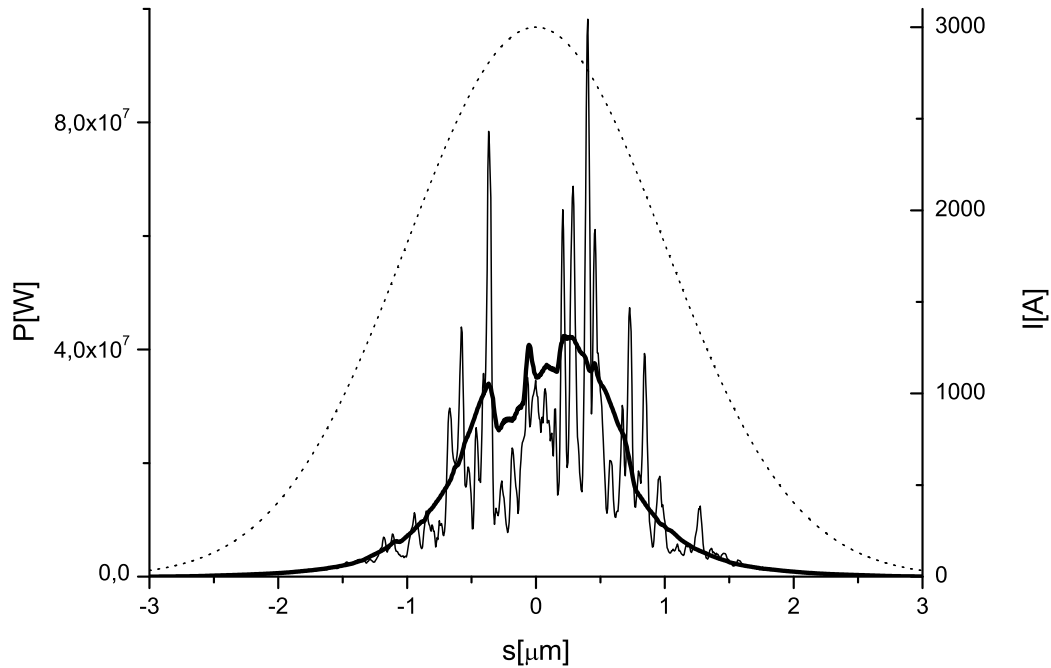}
\caption{Short pulse mode of operation. Input power at the crystal, at the end of the first undulator, $7$ cells long ($42$ m). The average input power is represented with a solid thick line. A typical shot is also shown with a solid thin line. The dashed line illustrates the corresponding distribution of the electron beam current.} \label{IPwin}
\end{figure}

After the first seven cells (42 m) the electron bunch is sent through the weak chicane, while radiation is filtered through a single diamond crystal. Here we use the C(400) Bragg reflection and we assume, as said before, that the crystal has a thickness of $0.1$ mm. The modulus and phase of the transmissivity function for the sigma-polarization component are shown in Fig. \ref{ImodT} and Fig. \ref{IphT}, and were calculated as described in \cite{OURX}. The profile of the radiation before the filter is shown in Fig. \ref{IPwin}.

\begin{figure}[tb]
\includegraphics[width=1.0\textwidth]{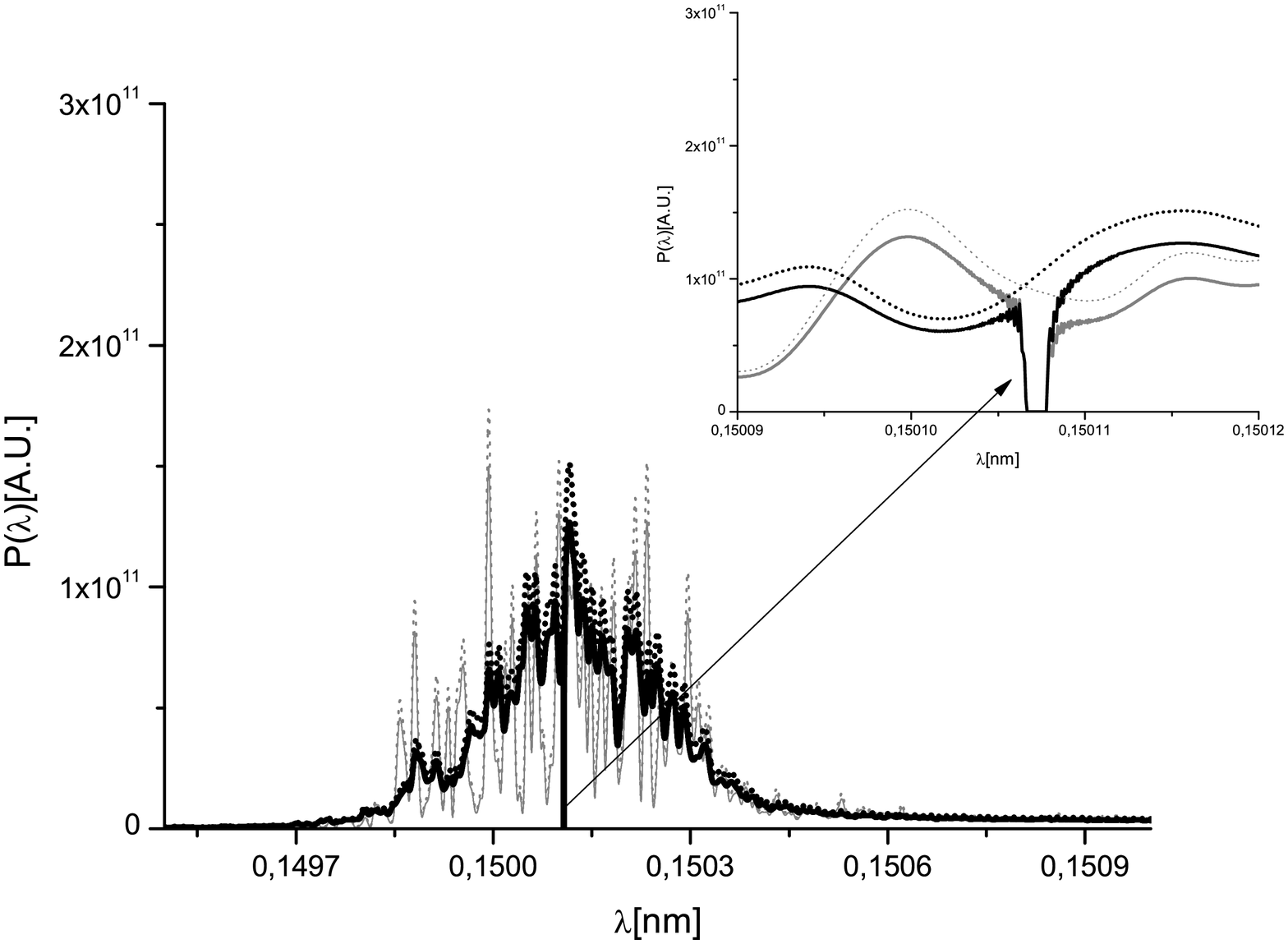}
\caption{Short pulse mode of operation. Average output spectrum after the diamond crystal and typical shot (solid thick line and solid thin line respectively). The bandstop effect is clearly visible, and highlighted in the inset. For comparison, the average spectrum before the diamond crystal and a typical shot (dotted thick line and dotted thin line respectively) is also shown.} \label{ISpcomp}
\end{figure}
\begin{figure}[tb]
\includegraphics[width=1.0\textwidth]{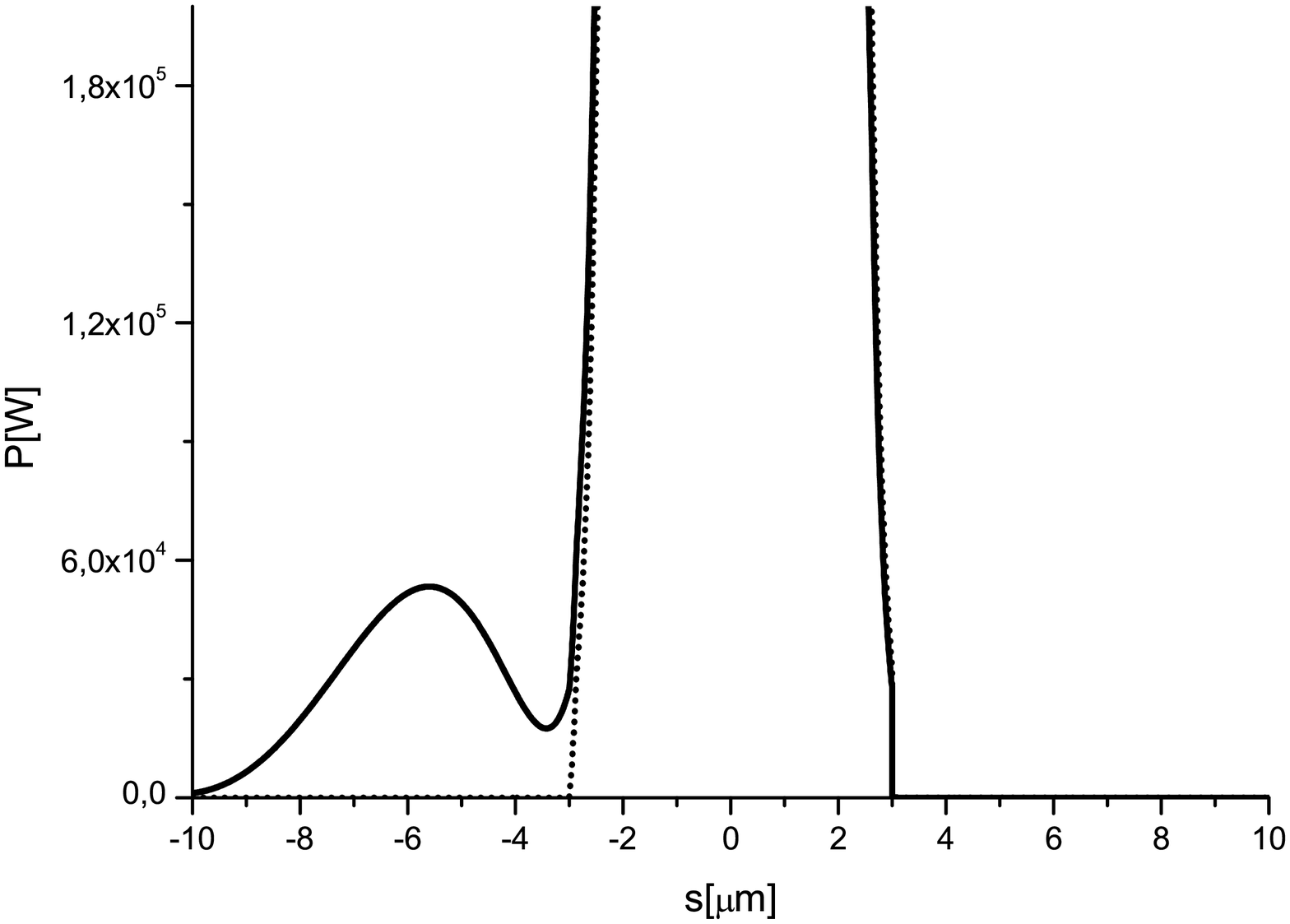}
\caption{Short pulse mode of operation. Power distribution before
(dotted line) and after (solid line) transmission through the
crystal. The monochromatic tail due to the transmission through the bandstop filter is evident on the left of the figure.} \label{IPcomp}
\end{figure}

As discussed above, and explained in detail in \cite{OURX}, the crystal acts as a bandstop filter. The effect is best shown by a comparison of the spectra before and after the filter, Fig. \ref{ISpcomp}.  Monochromatization does not take place in the frequency domain. At first glance, the passage through the bandstop filter is only responsible for slight change in the power distribution along the pulse. However, a zoom of the vertical axis shows what we are interested in: a long, monochromatic tail in the power distribution on the left side of the picture, Fig. \ref{IPcomp}.

Following the first crystal, we consider two alternative schemes. First, a two-undulator configuration, where the radiation from the first crystal is used to seed the electron bunch in an output undulator and, second, a three-undulator configuration where two monochromatization cascades are foreseen. Let us first consider the two-undulator configuration.

\begin{figure}[tb]
\includegraphics[width=1.0\textwidth]{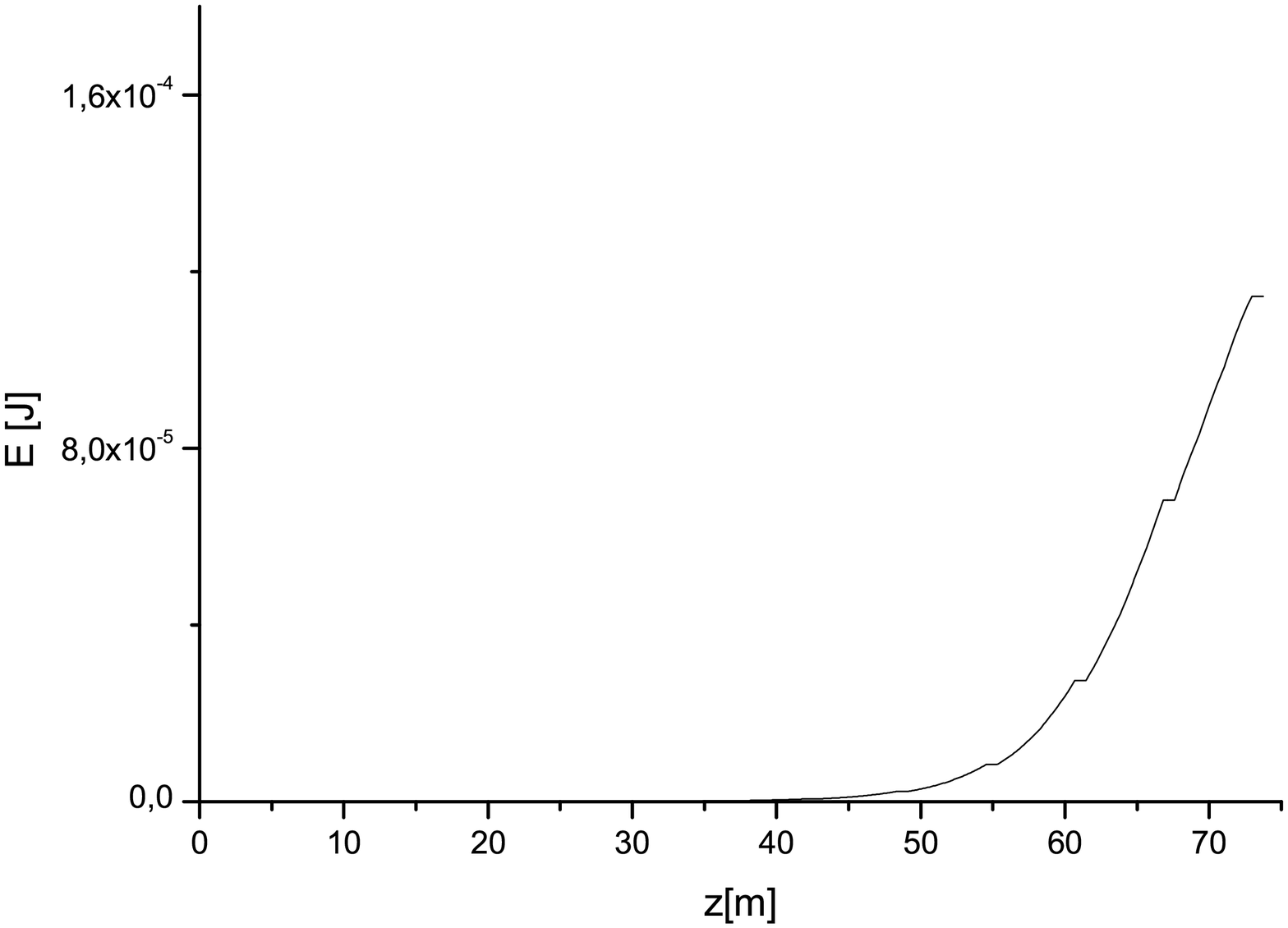}
\caption{Two-undulator configuration. Short pulse mode of operation. Average energy of the radiation pulse as a function of the second undulator length.} \label{EN2}
\end{figure}
\begin{figure}[tb]
\includegraphics[width=1.0\textwidth]{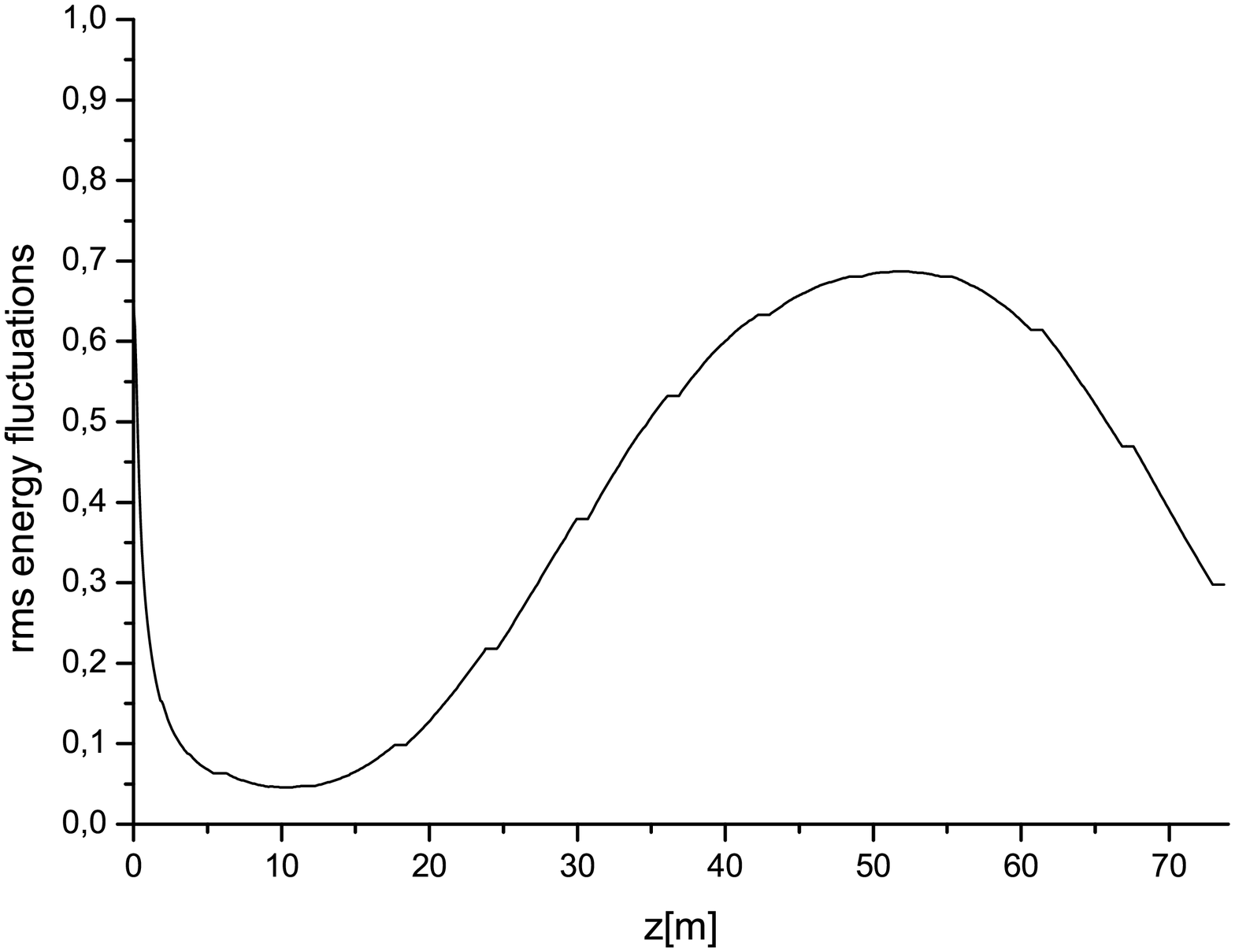}
\caption{Two-undulator configuration. Short pulse mode of operation. rms energy deviation from the average as a function of the second undulator length.} \label{RMS2}
\end{figure}

\begin{figure}[tb]
\includegraphics[width=1.0\textwidth]{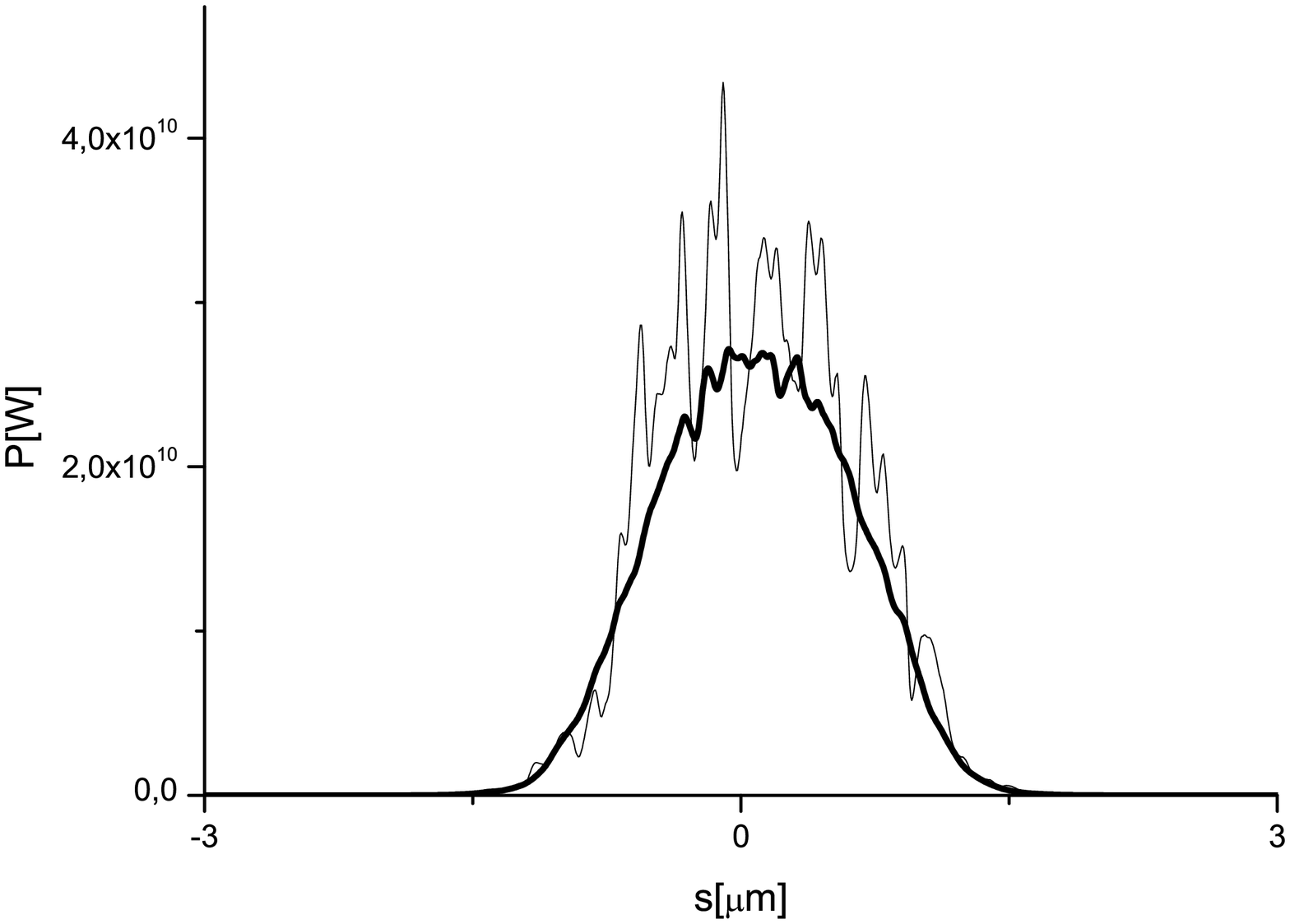}
\caption{Two-undulator configuration. Short pulse mode of operation. Average and typical single-shot output power (respectively, thick and thin solid lines). } \label{PW2}
\end{figure}

\begin{figure}[tb]
\includegraphics[width=1.0\textwidth]{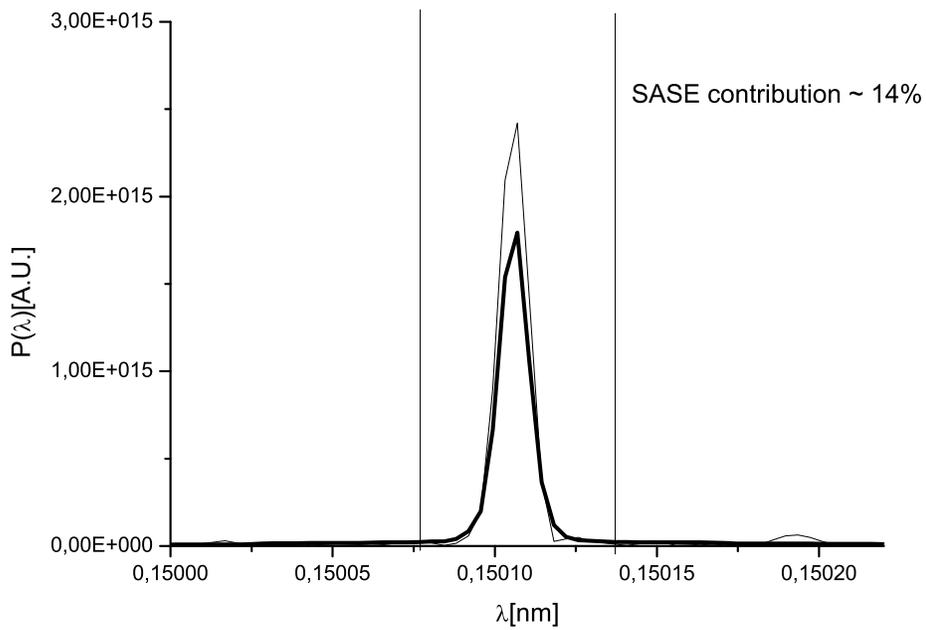}
\caption{Two-undulator configuration. Short pulse mode of operation. Average and typical single-shot output spectrum (respectively, thick and thin solid lines). An estimation of the SASE contribution can be done by evaluating the total power outside the spectral window shown with black straight lines and dividing it to the spectral power integrated over all the spectrum.} \label{Spout_2}
\end{figure}

With reference to Fig. \ref{cs6}, we optimized the length of the output undulator to obtain maximal spectral power. The optimal length is found to be $12$ cells. In Fig. \ref{EN2} and Fig. \ref{RMS2} we show, respectively, the average energy of the pulse as a function of the undulator length, and the rms deviation from the average\footnote{As one may see, at the beginning of the
undulator the fluctuations of energy per pulse apparently drop, see Fig. \ref{RMS2}. This can be explained considering the fact that the Genesis output consists of the total power integrated over the full grid up to an artificial boundary, i.e. there is no spectral selection. Therefore, our calculations include a relatively large spontaneous emission background, which has a much larger spectral width with respect to the amplification bandwidth and which fluctuates negligibly from shot to shot. Since there is a long lethargy of the seeded radiation at the beginning of the FEL amplifier, one observes an apparent decrease of fluctuations. Then, when lethargy ends, the seed pulse gets amplified and fluctuations effectively return to about the same level as after monochromator.}. Fig. \ref{PW2} and Fig. \ref{Spout_2} show, instead, the output power and spectrum. An estimation of the SASE contribution can be done by evaluating the total power outside the spectral window shown with black straight lines in Fig. \ref{Spout_2} and dividing it by the total power in the pulse. Such ratio is about $14\%$.  Obviously there is some ambiguity in the definition of the spectral window. Nevertheless, the ratio is weakly dependent on the choice of the window width and one can take the over-mentioned figure as a rough estimation of the SASE contribution to the total pulse.

\clearpage

\subsection{Three-undulator configuration}

\begin{figure}[tb]
\includegraphics[width=1.0\textwidth]{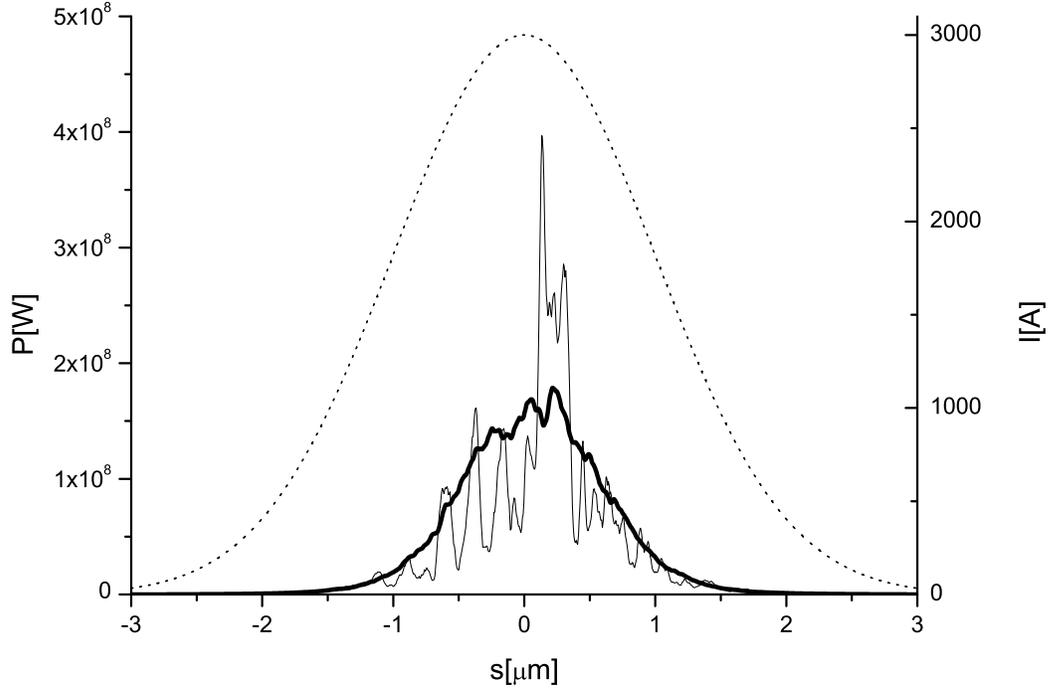}
\caption{Short pulse mode of operation, second monochromatization cascade. Input power at the second crystal, at the end
of the second undulator, $7$ cells long ($42$ m). The average input power is represented with a solid thick line. A typical shot is also shown with a solid thin line. The dashed line illustrates the corresponding distribution of the electron beam current.} \label{IIPwin}
\end{figure}

As discussed before, the three-undulator configuration presents advantages related to the high contrast between seeded and SASE signal. With reference to Fig. \ref{cs2} the second undulator is now shortened to $7$ cells (corresponding to $42$ m), and followed by a second seeding stage. The input power impinging on the second crystal is shown in Fig. \ref{IIPwin}.

\begin{figure}[tb]
\includegraphics[width=1.0\textwidth]{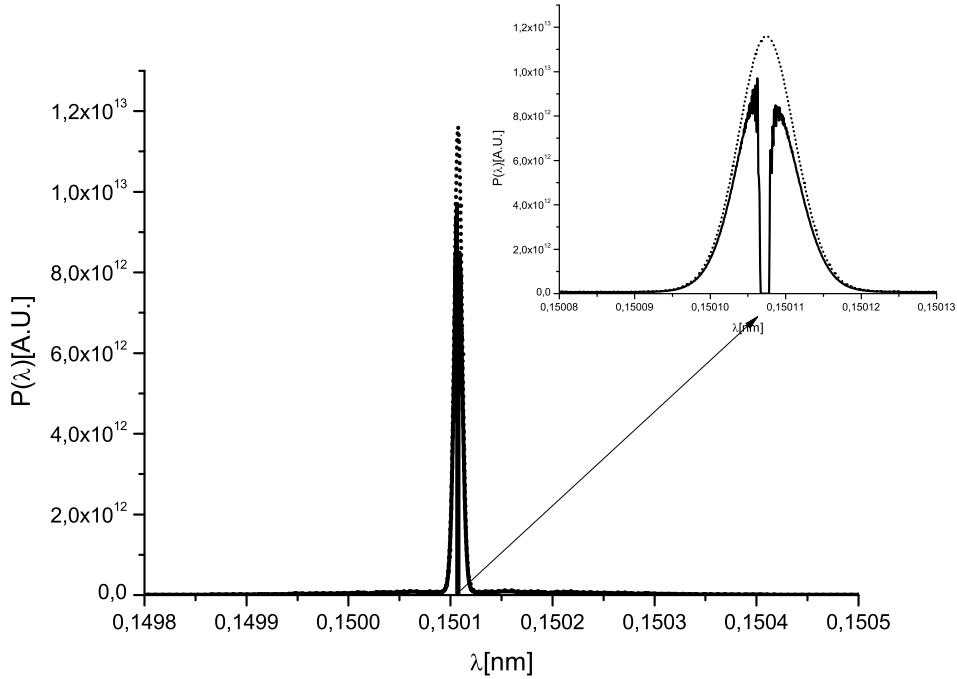}
\caption{Short pulse mode of operation, second monochromatization cascade. Average output spectrum after the second diamond crystal (solid thick line). The bandstop effect is clearly visible, and highlighted in the inset. For comparison, the average spectrum before the diamond crystal (dotted thick line) is also shown.} \label{IISpcomp}
\end{figure}
\begin{figure}[tb]
\includegraphics[width=1.0\textwidth]{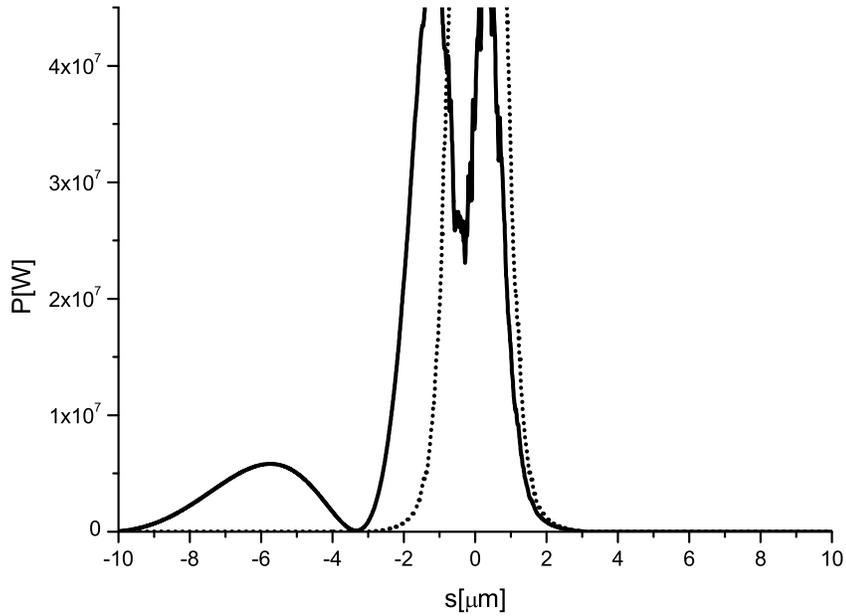}
\caption{Short pulse mode of operation, second monochromatization cascade. Power distribution before (dotted line) and after (solid line) transmission through the second
crystal. The monochromatic tail due to the transmission through the bandstop filter is evident on the left of the figure.} \label{IIPcomp}
\end{figure}

As before, the effect of the filter is best shown by a comparison of the spectra before and after the filter, Fig. \ref{IISpcomp}.  A long, monochromatic tail in the power distribution on the left side of Fig. \ref{IIPcomp} constitutes the time-domain effect of the filtering procedure.

It is interesting to discuss here a paradox which arises when one compares the average spectrum before the filter (dotted line in Fig. \ref{IISpcomp}) and the average spectrum after the filter in the first monochromatization cascade, Fig. \ref{ISpcomp}. The bandstop filter effect is visible in the inset of the latter figure, but after amplification (dotted line in Fig. \ref{IISpcomp}) it disappears, and no hole is visible in the spectrum anymore. This leads to an apparent contradiction, because the radiation pulse simply passes through a linear amplifier, and the bandstop effect should still be visible. The paradox is explained in terms of the windowing process. The monochromatic pulse at the exit of the first filter serves, in fact, as seed for the electron bunch, which is shorter than the monochromatic wake itself. The seeding field is effectively sampled only within the electron bunch. In other words, the temporal windowing process operated by the electron bunch is equivalent to a spectral measurement over a time equal to the bunch duration $\sigma_e/c$. The length of the monochromatic tail of the electric field (see Fig. \ref{IPcomp}) is related to the bandwidth of the crystal by inverse proportionality and is about $1/\Delta \omega_m$. When the two bandwidths are related by $\sigma_e/c \ll 1/\Delta \omega_m$, like in the short-pulse mode of operation, the windowing process in the time domain is equivalent to a convolution, in the frequency domain, of the seeding pulse and a window-like signal with bandwith $c/\sigma_e \gg \Delta \omega_m$. It follows that the hole in the spectrum is not resolved or, in other words, that we are measuring the spectrum of the seed signal with an instrument (the electron bunch) which integrates over a temporal interval too short to obtain the proper resolution. The situation changes when the bunch length increases. This is equivalent to an increase in resolution in the spectral measurement, and can be seen in Fig. \ref{IISpcomp_L}, which is the analogous of Fig. \ref{IISpcomp} for the long-pulse mode of operation. The electron bunch is not long enough to resolve the seed signal properly in the frequency domain, but the first filtering process is still visible in the dotted spectrum before the second filter, due to a better spectral resolution.

%

%

\begin{figure}[tb]
\includegraphics[width=1.0\textwidth]{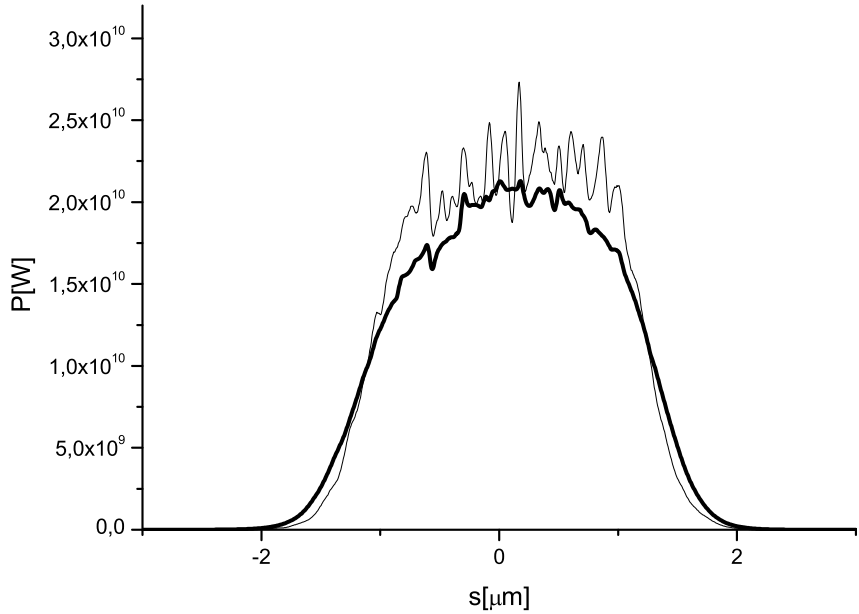}
\caption{Three-undulator configuration. Short pulse mode of operation. Average and typical single-shot output power (respectively, thick and thin solid lines).} \label{Pout}
\end{figure}

\begin{figure}[tb]
\includegraphics[width=1.0\textwidth]{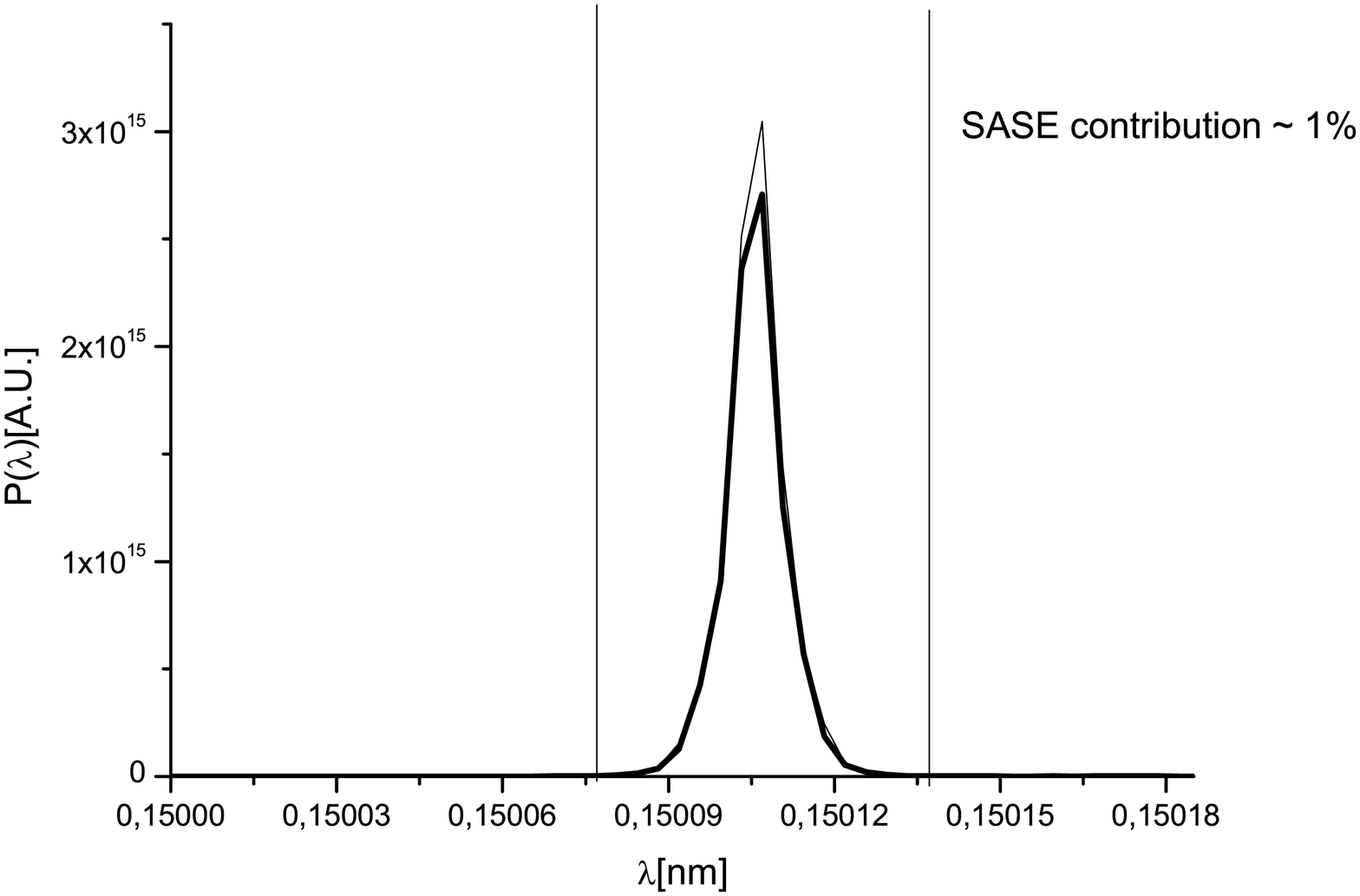}
\caption{Three-undulator configuration. Short pulse mode of operation. Average and typical single-shot output spectrum (respectively, thick and thin solid lines). An estimation of the SASE contribution can be done by evaluating the total power outside the spectral window shown with black straight lines and dividing it to the spectral power integrated over all the spectrum.} \label{Spout}
\end{figure}

\begin{figure}[tb]
\includegraphics[width=1.0\textwidth]{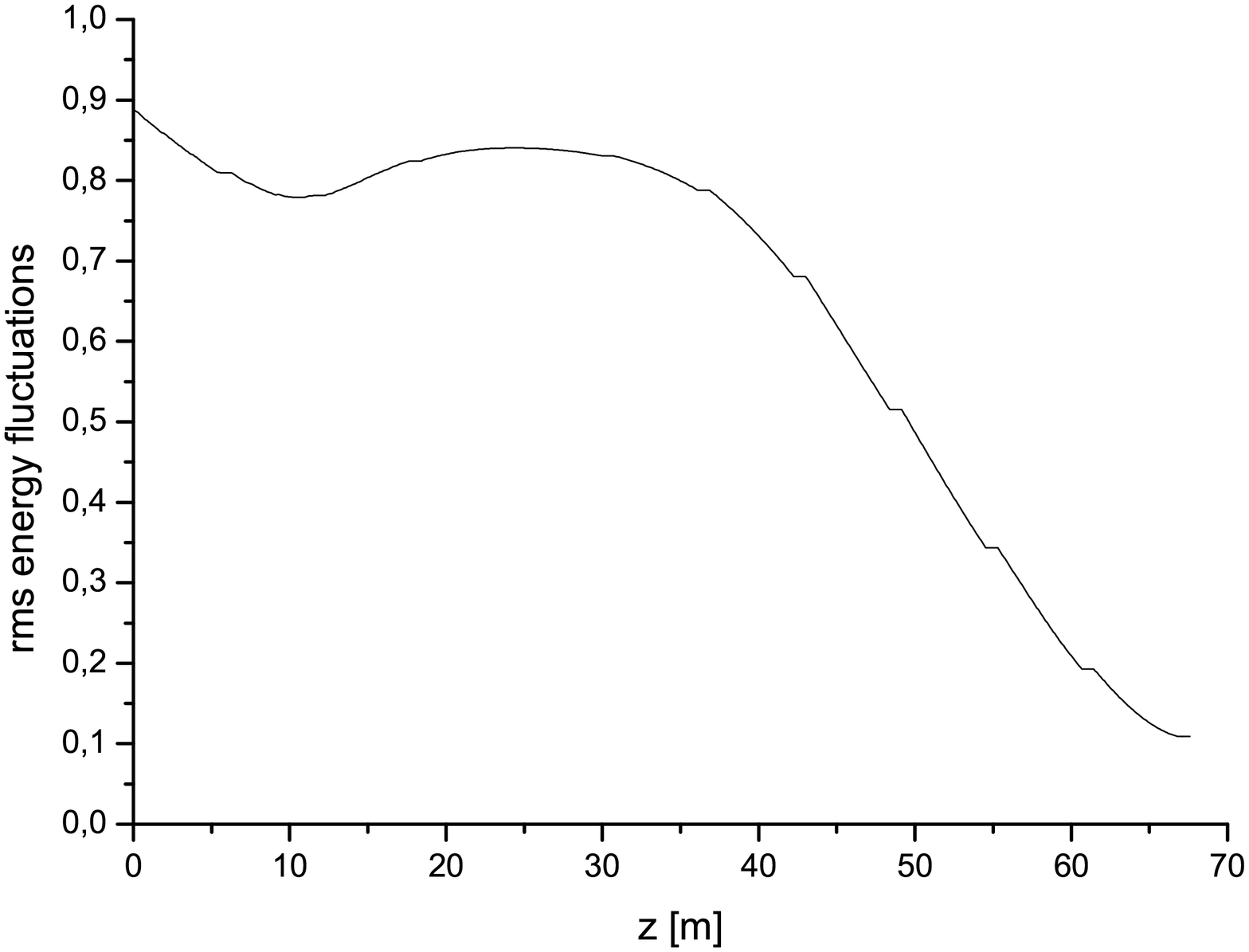}
\caption{Three-undulator configuration. Short pulse mode of operation. rms energy deviation from the average as a function of the third undulator length.} \label{Srms}
\end{figure}

\begin{figure}[tb]
\includegraphics[width=1.0\textwidth]{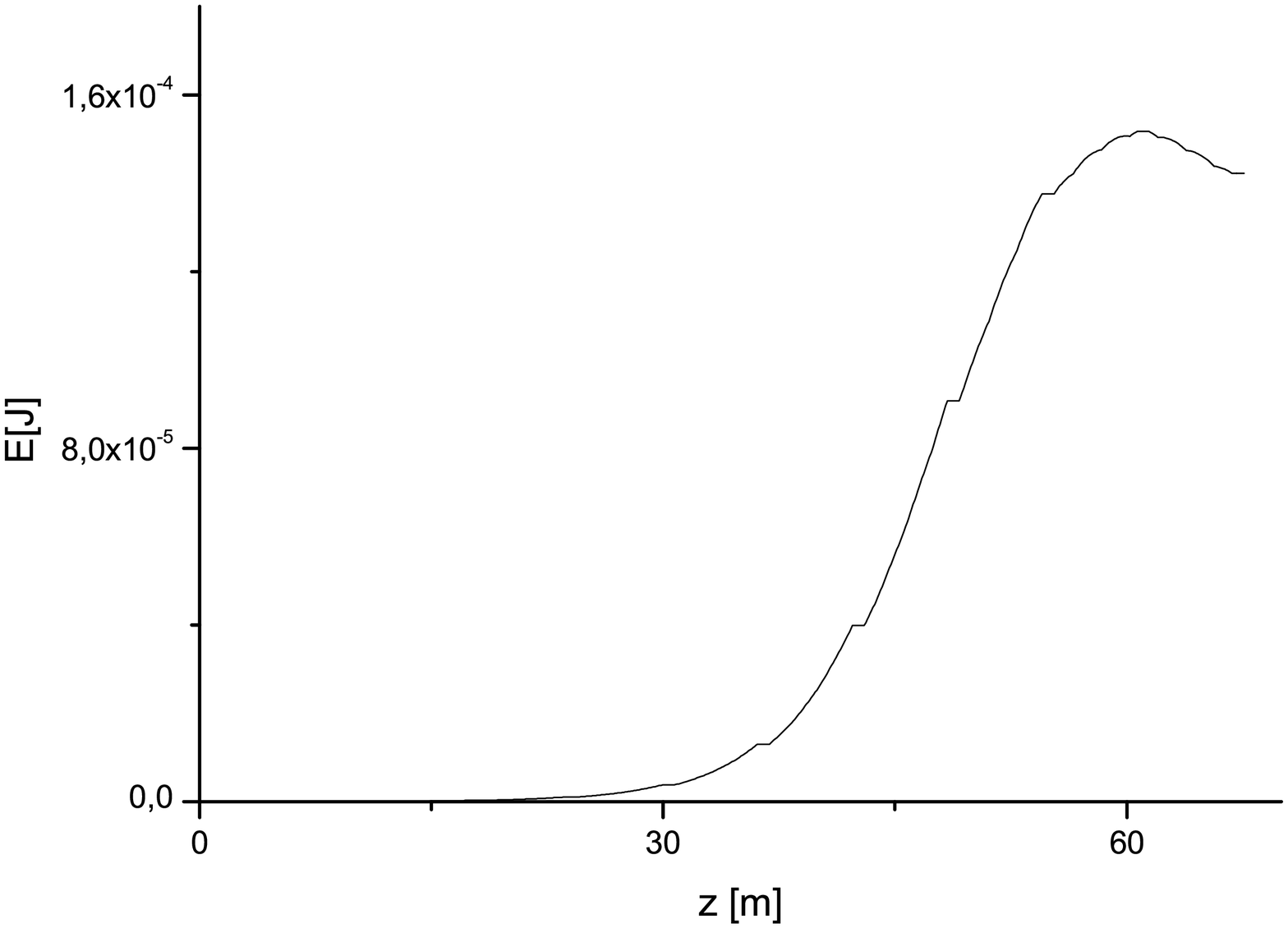}
\caption{Three-undulator configuration.  Short pulse mode of operation. Average energy of the radiation pulse as a function of the third undulator length.} \label{Senergy}
\end{figure}

Following the second crystal, the radiation is used to seed once more the electron bunch. Radiation is collected at the exit of a third undulator. Fig. \ref{Pout} and Fig. \ref{Spout} respectively show the output power and spectrum for the three-undulator configuration, while in Fig. \ref{Srms} and Fig. \ref{Senergy} we present, respectively, the rms deviation from the average and the average energy of the pulse as a function of the undulator length. Similarly as before, an estimation of the SASE contribution can be done by evaluating the total power outside the spectral window shown with black straight lines in Fig. \ref{Spout}, and dividing it by the spectral power integrated over all the spectrum yielding, as expected, a much smaller SASE contribution in the order of a percent.


\section{\label{Feal} Feasibility study for the long-bunch mode of operation}

Let us now consider the case of long pulse mode of operation. The setups for the two-undulator configuration and for the three-undulator configuration are the same considered before, respectively in Fig. \ref{cs6} and Fig. \ref{cs2}. As said above, for the long pulse
mode of operations Table \ref{tt1} is still valid, except for a
ten times larger charge ($0.25$ nC) and a ten times longer rms
bunch length ($10 \mu$m).

\subsection{Two-undulator configuration}

\begin{figure}[tb]
\includegraphics[width=1.0\textwidth]{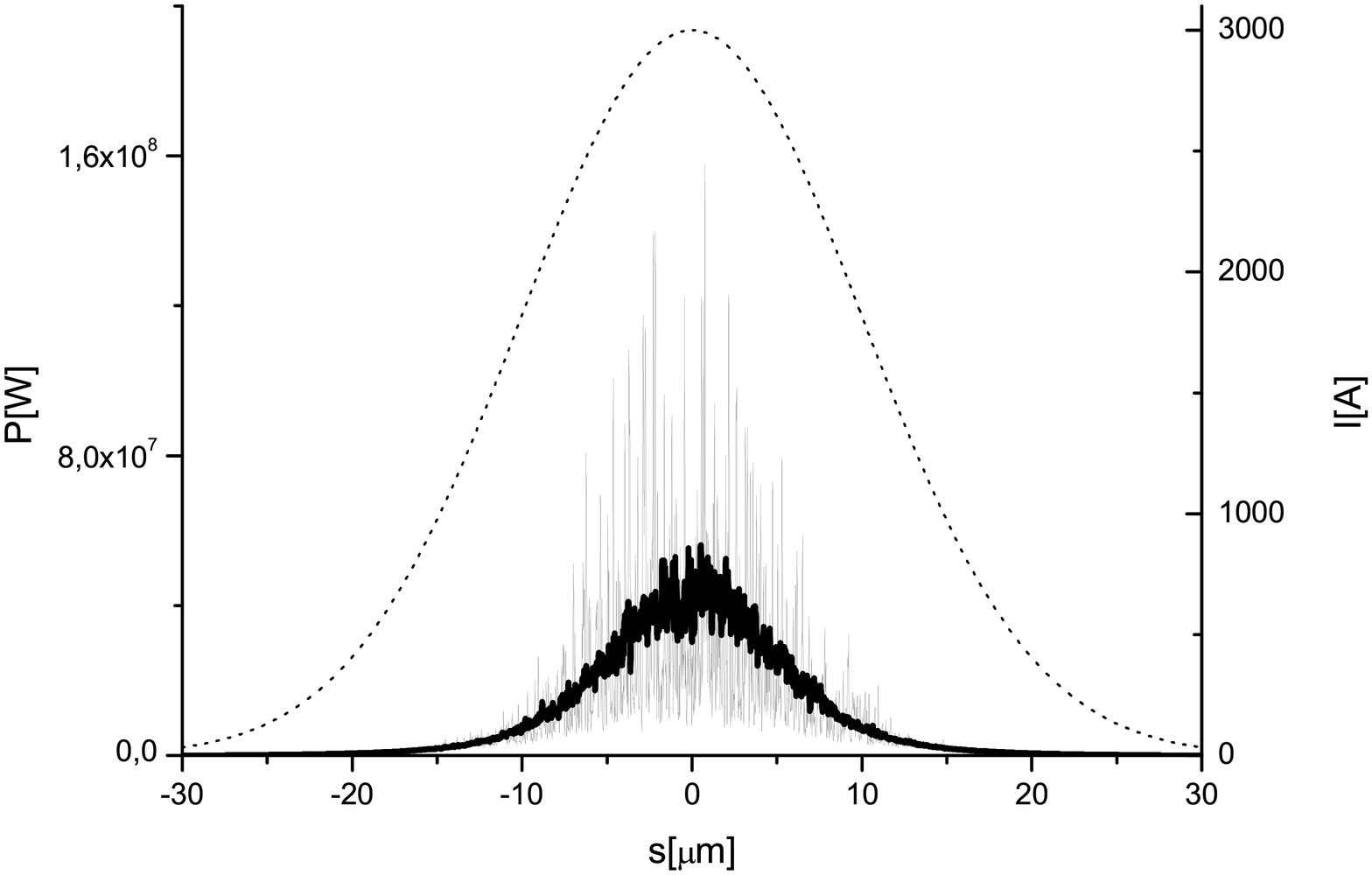}
\caption{Long pulse mode of operation. Input power at the crystal, at the end of the first undulator, $7$ cells long ($42$ m). The average input power is represented with a solid thick line. A typical shot is also shown with a solid thin line. The dashed line illustrates the corresponding distribution of the electron beam current.} \label{IPwin_L}
\end{figure}

Similarly as before, we begin with the first monochromatization cascade. After the first seven cells (42 m) the electron bunch is sent through the weak chicane, while radiation is filtered through a single diamond crystal. We use the same crystal as before. The transmissivity function (modulus and phase) for the sigma-polarization component was already presented in Fig. \ref{ImodT}, and Fig.\ref{IphT}. The profile of the radiation before the filter is shown in Fig. \ref{IPwin_L}.

\begin{figure}[tb]
\includegraphics[width=1.0\textwidth]{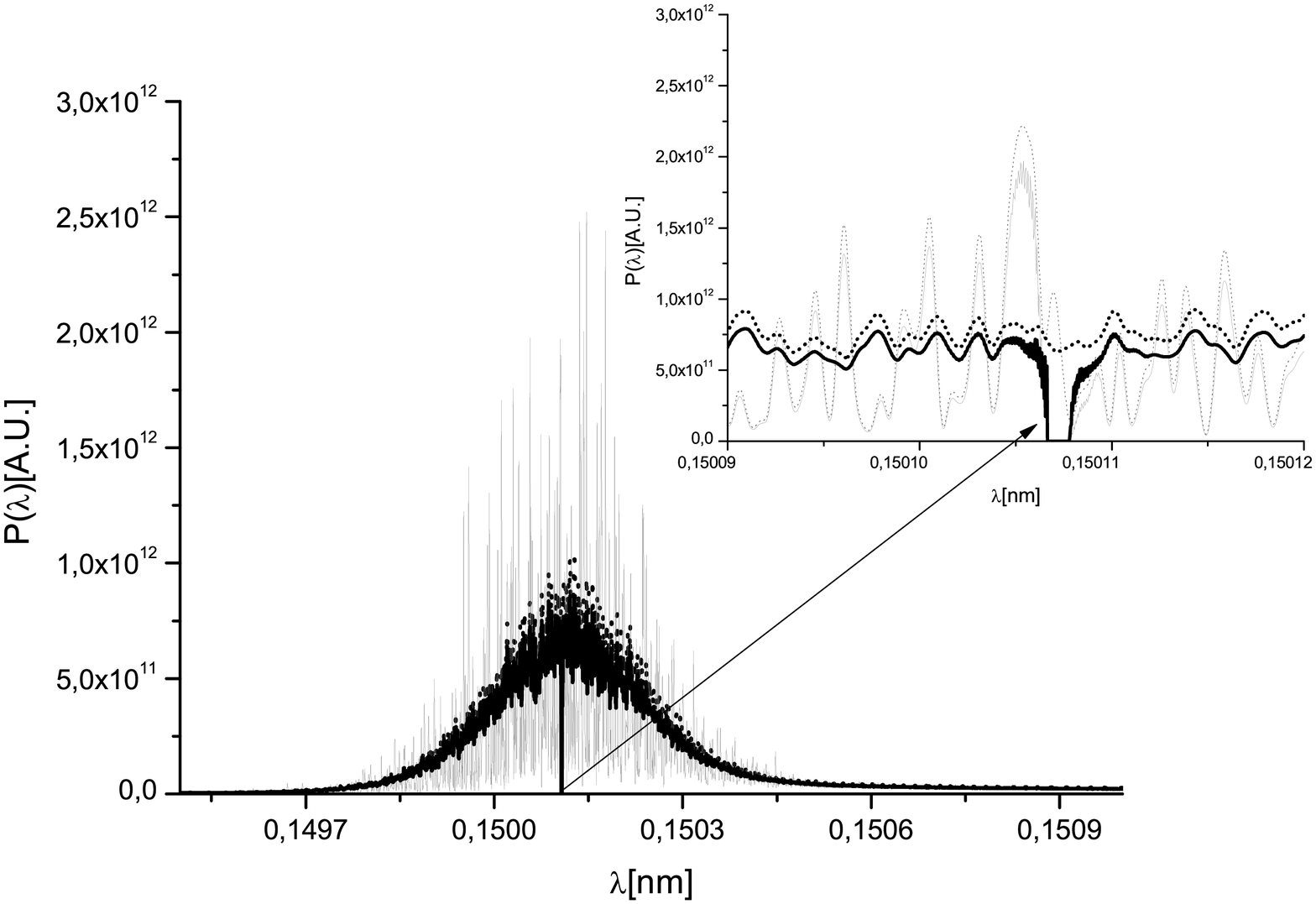}
\caption{Long pulse mode of operation. Average output spectrum after the diamond crystal and typical shot (solid thick line and solid thin line respectively). The bandstop effect is clearly visible, and highlighted in the inset. For comparison, the average spectrum before the diamond crystal and a typical shot (dotted thick line and dotted thin line respectively) is also shown.} \label{ISpcomp_L}
\end{figure}
%

%

\begin{figure}[tb]
\includegraphics[width=1.0\textwidth]{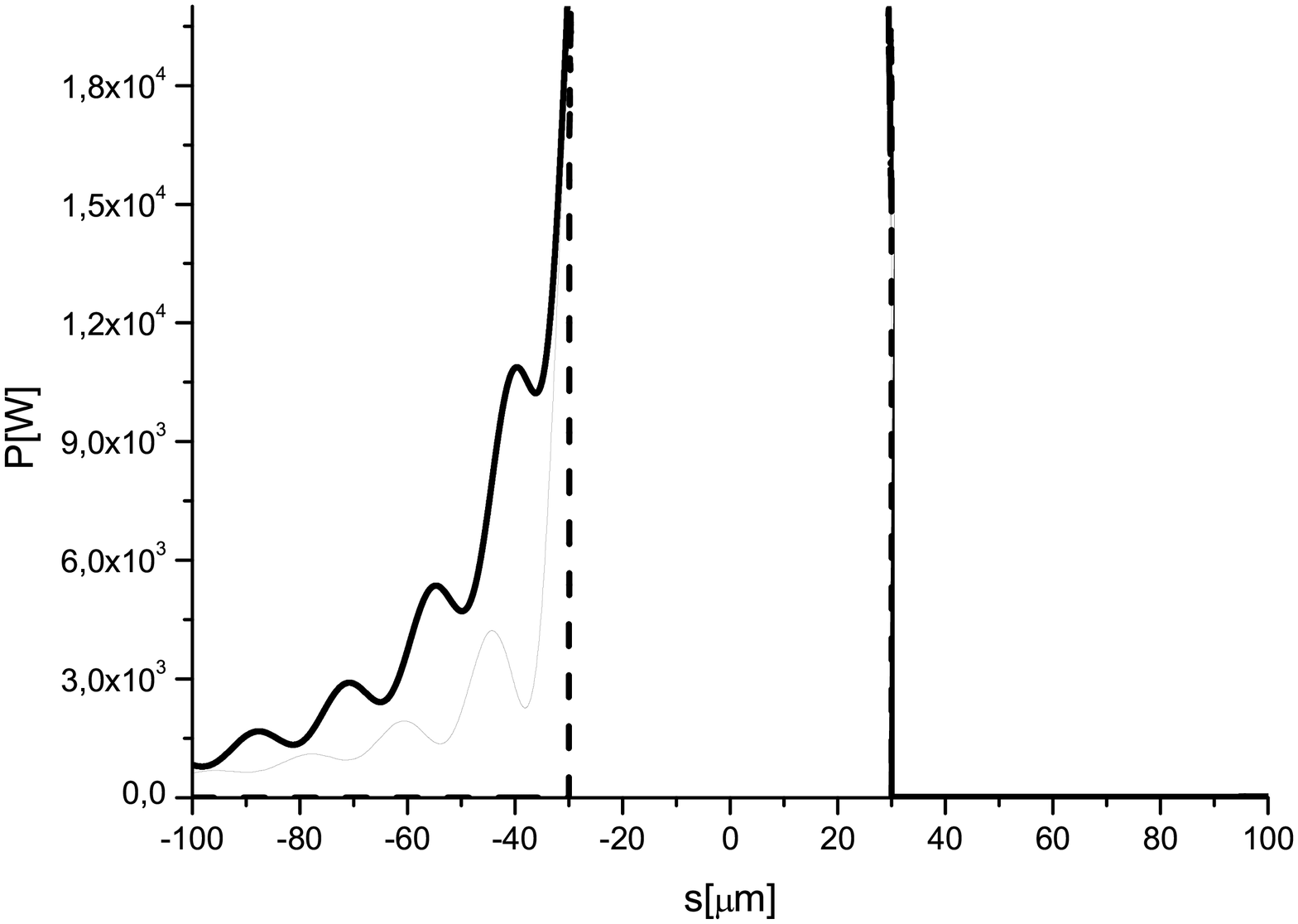}
\caption{Long pulse mode of operation. Power distribution before
(dotted line) and after (solid line) transmission through the
crystal. The monochromatic tail due to the transmission through the bandstop filter is evident on the left of the figure.} \label{IPcomp_L}
\end{figure}
The effect is best shown by a comparison of the spectra before and after the filter, Fig. \ref{ISpcomp_L}.  A long, monochromatic tail in the power distribution can be clearly seen on the left side of  Fig. \ref{IPcomp_L}. Compared to the previously discussed short pulse mode of operation, the seeding level is about an order of magnitude smaller.

Similarly as before, following the first crystal, we consider, first, a two-undulator configuration and, second, a three-undulator configuration. Although the average power of the seed signal after the first crystal amounts only to about $3$ kW, the bunch length is now $10$ times longer. Therefore, the contrast in spectrum can be comparable with the two-undulator configuration in the short pulse scheme.

\begin{figure}[tb]
\includegraphics[width=1.0\textwidth]{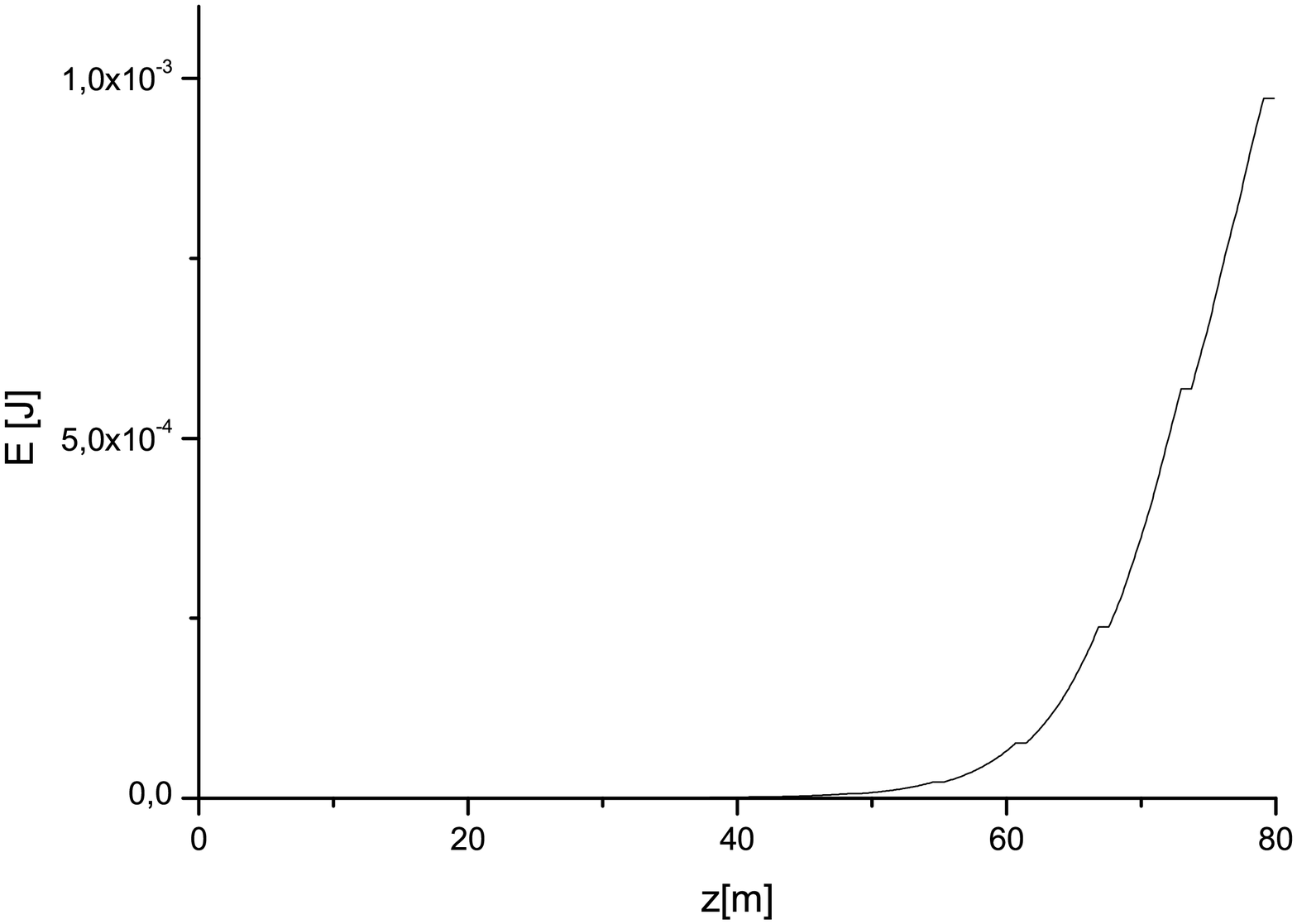}
\caption{Two-undulator configuration. Long pulse mode of operation. Average energy of the radiation pulse as a function of the second undulator length.} \label{EN2L}
\end{figure}
\begin{figure}[tb]
\includegraphics[width=1.0\textwidth]{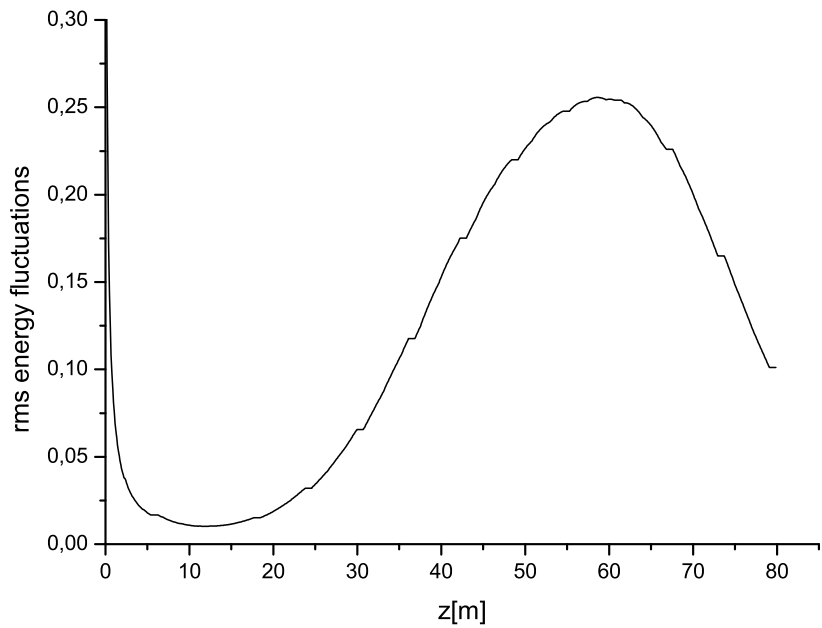}
\caption{Two-undulator configuration. Long
pulse mode of operation. RMS pulse energy fluctuations along the second undulator.} \label{RMS2L}
\end{figure}
\begin{figure}[tb]
\includegraphics[width=1.0\textwidth]{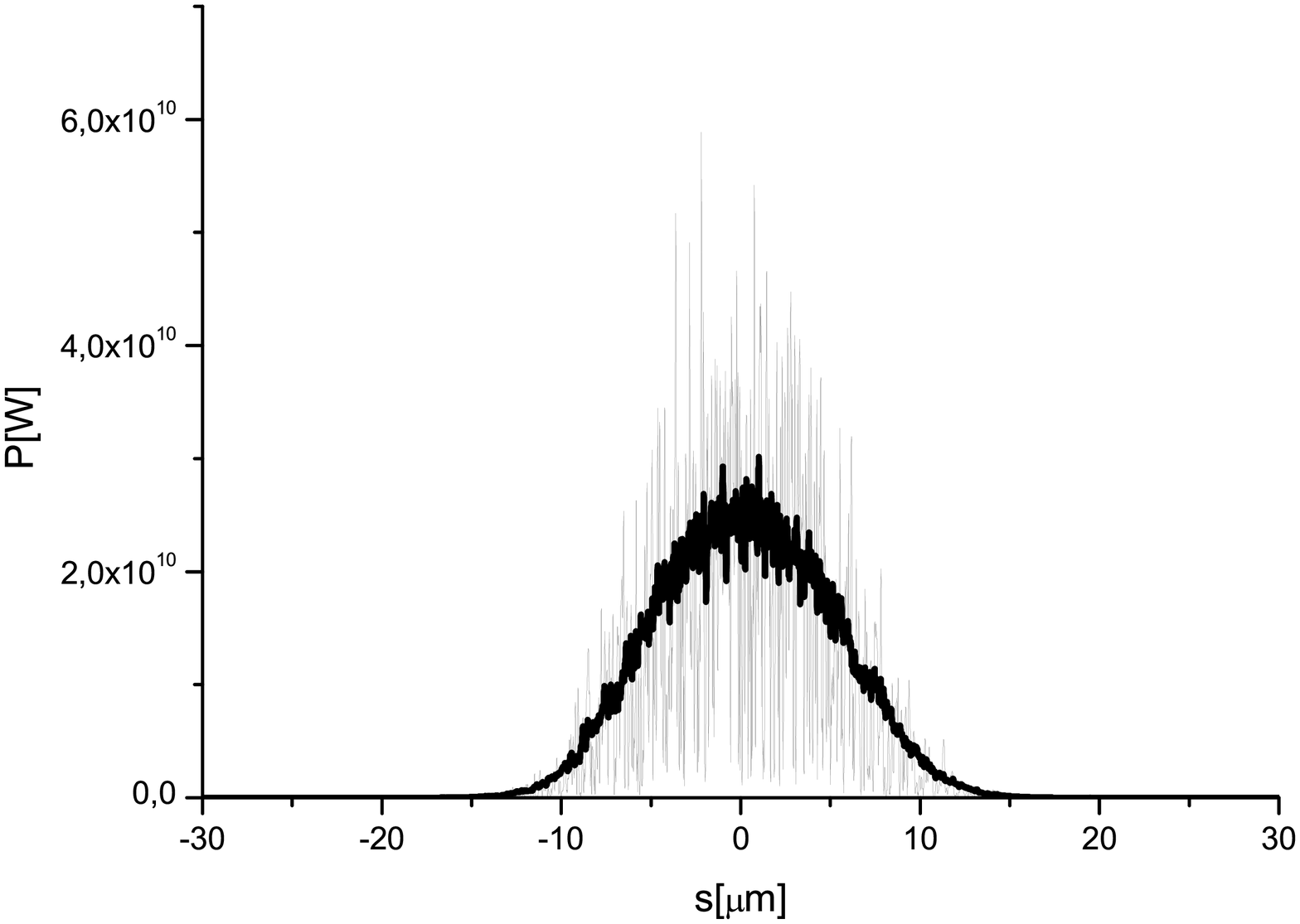}
\caption{Two-undulator configuration. Long pulse mode of operation. Average and typical single-shot output spectrum (respectively, thick and thin solid lines).} \label{PW2L}
\end{figure}

\begin{figure}[tb]
\includegraphics[width=1.0\textwidth]{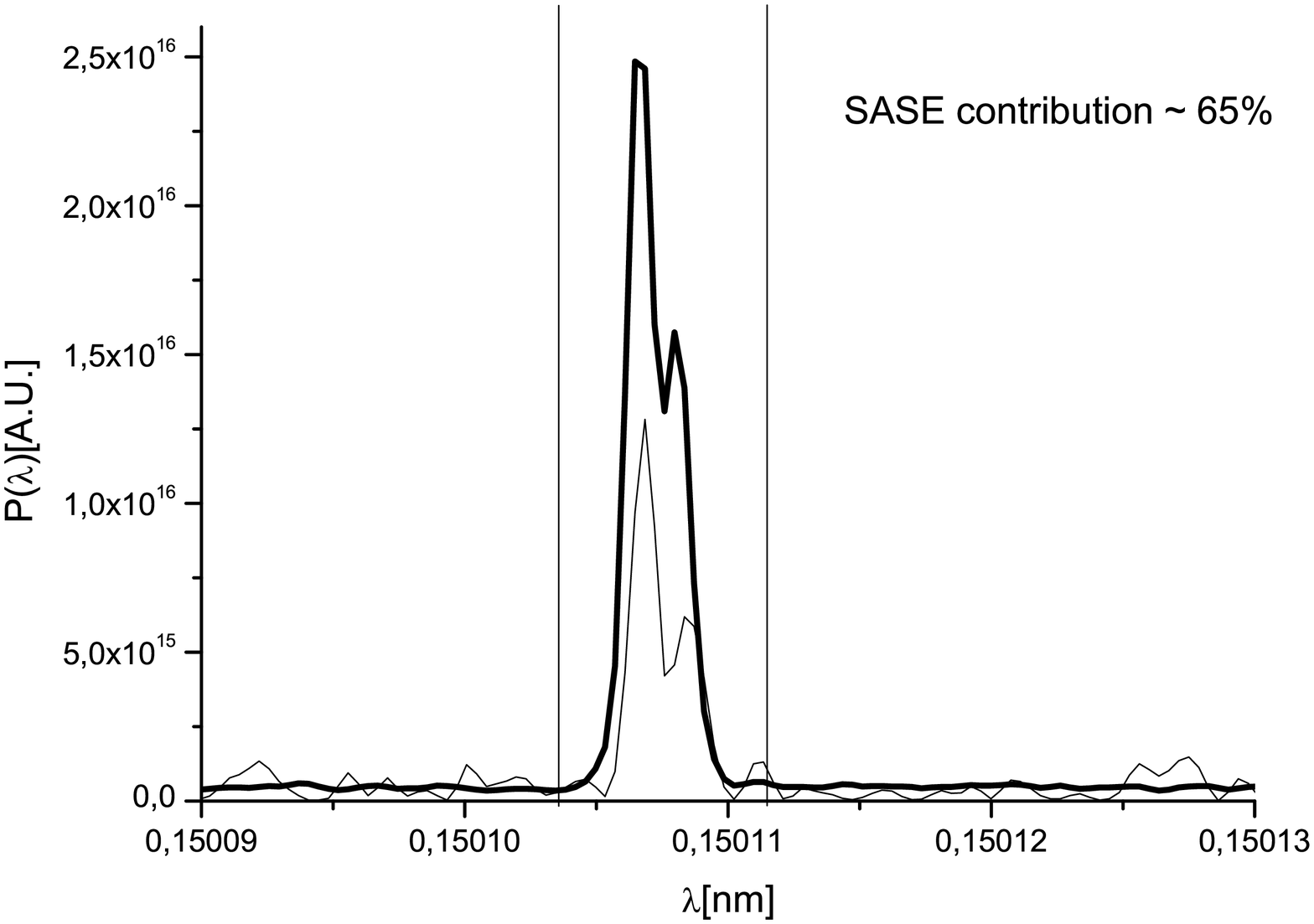}
\caption{Two-undulator configuration. Long pulse mode of operation. Average and typical single-shot output spectrum (respectively, thick and thin solid lines). An estimation of the SASE contribution can be done by evaluating the total power outside the spectral window shown with black straight lines and dividing it to the spectral power integrated over all the spectrum.} \label{Spout_2L}
\end{figure}

Still with reference to Fig. \ref{cs6}, we optimized the length of the output undulator to obtain maximal spectral power. As before, the optimal length is found to be $12$ cells. In Fig. \ref{EN2L} and Fig. \ref{RMS2L} we show, respectively, the average energy of the pulse as a function of the undulator length, and the rms deviation from the average. Fig. \ref{PW2L} and Fig. \ref{Spout_2L} show, instead, the output power and spectrum. In the present case, an estimation of the SASE contribution yields an important SASE contribution in the order of $65\%$.

As before, we should remark that the two-undulator configuration may be particularly advantageous in cases when the total available undulator length is too short to enforce the three-undulator configuration. In this case, one may consider an increase in length of the first undulator in order to increase the contrast between seeded and SASE signal, at the cost of increasing the heat load.

\clearpage

\subsection{Three-undulator configuration}

\begin{figure}[tb]
\includegraphics[width=1.0\textwidth]{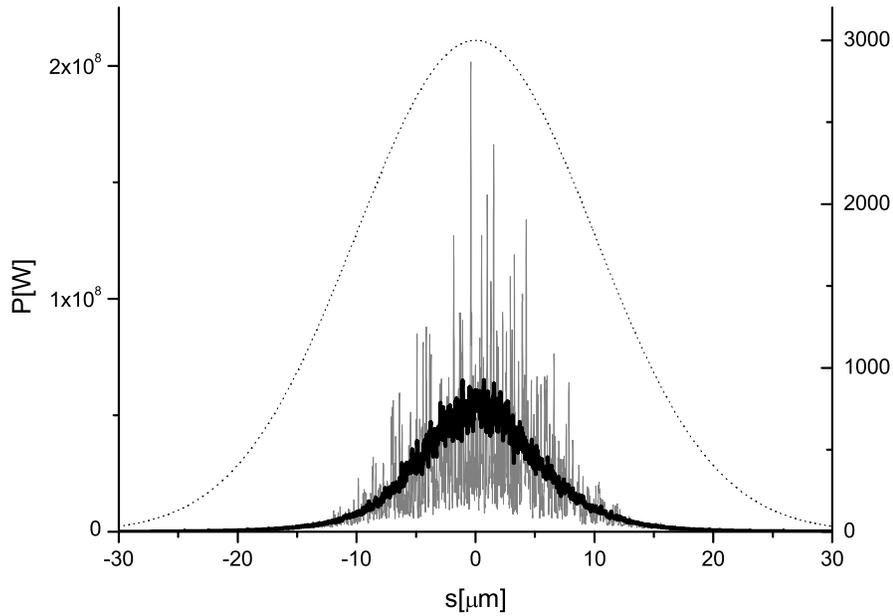}
\caption{Long pulse mode of operation, second monochromatization cascade. Input power at the second crystal, at the end
of the second undulator, $7$ cells long ($42$ m). The average input power is represented with a solid thick line. A typical shot is also shown with a solid thin line. The dashed line illustrates the corresponding distribution of the electron beam current.} \label{IIPwin_L}
\end{figure}

As in the short pulse mode of operation, the three-undulator configuration presents advantages related to the high contrast between seeded and SASE signal. With reference to Fig. \ref{cs2} the second undulator is now shortened to $7$ cells (corresponding to $42$ m), and followed by a second seeding stage. The input power impinging on the second crystal is shown in Fig. \ref{IIPwin_L}.

\begin{figure}[tb]
\includegraphics[width=1.0\textwidth]{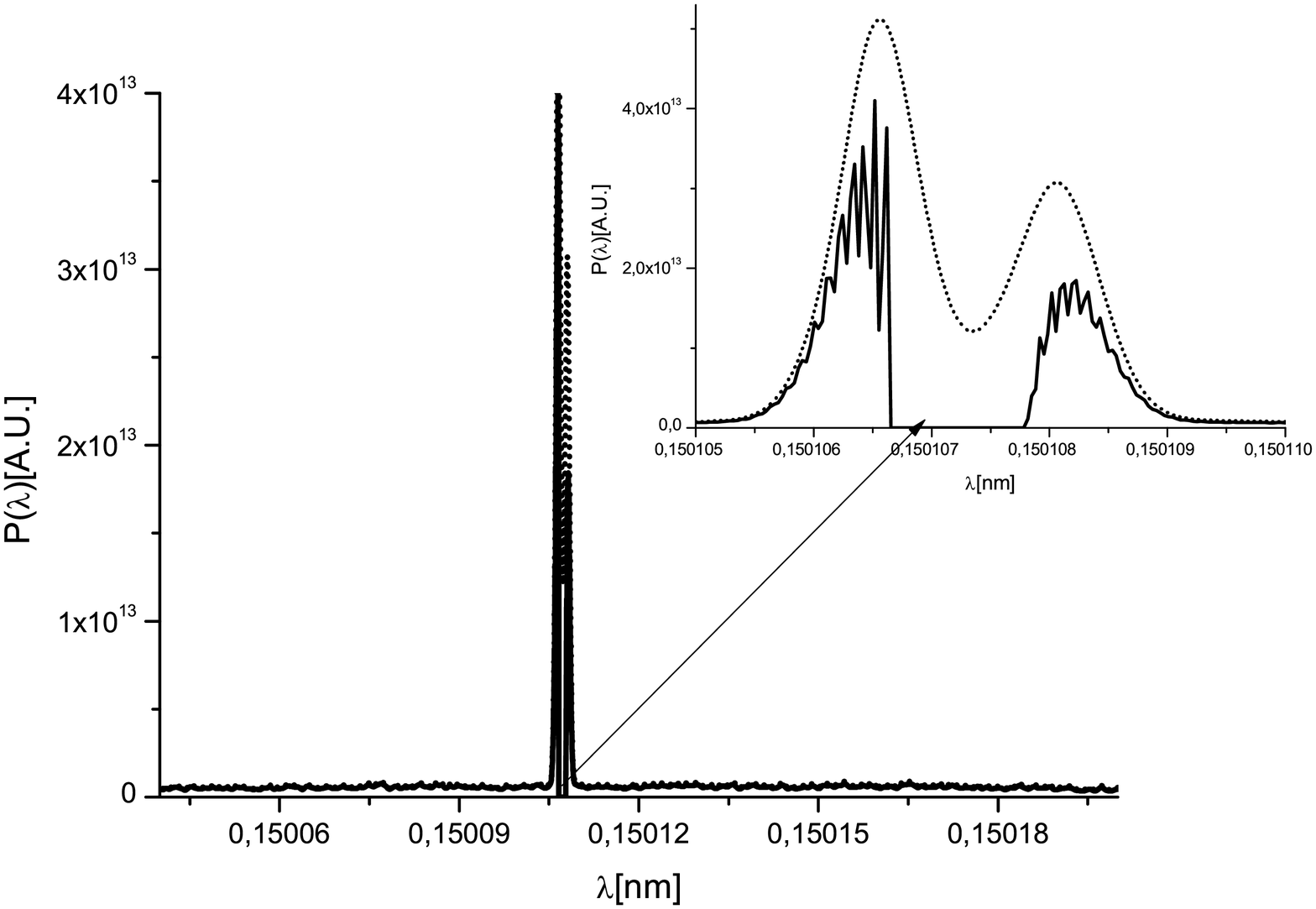}
\caption{Long pulse mode of operation, second monochromatization cascade. Average output spectrum after the second diamond crystal  (solid thick line). The bandstop effect is clearly visible, and highlighted in the inset. For comparison, the average spectrum before the diamond crystal (dotted thick line) is also shown.} \label{IISpcomp_L}
\end{figure}

\begin{figure}[tb]
\includegraphics[width=1.0\textwidth]{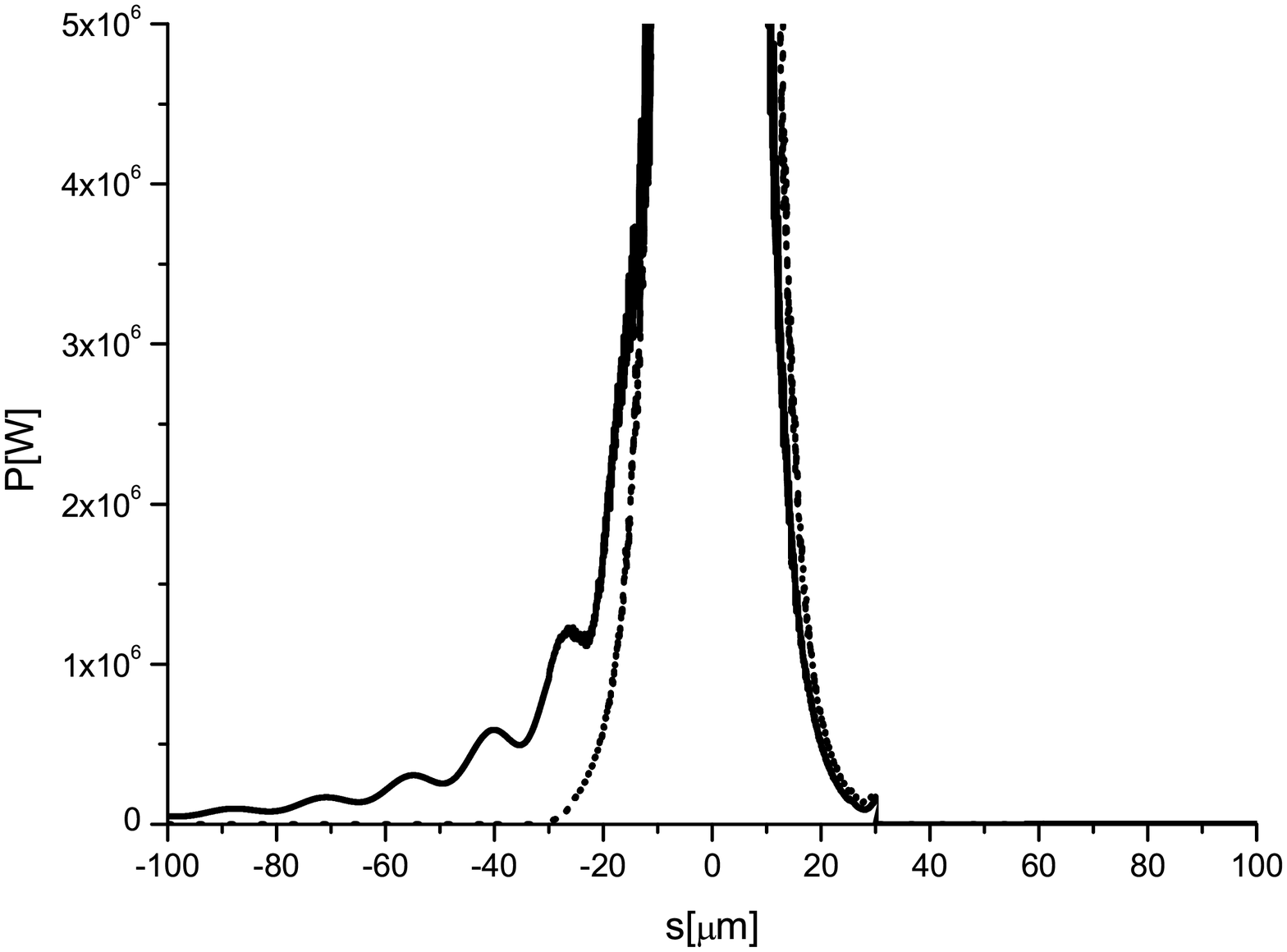}
\caption{Long pulse mode of operation, second monochromatization cascade. Power distribution before (dotted line) and after (solid line) transmission through the second
crystal. The monochromatic tail due to the transmission through the bandstop
filter is evident on the left of the figure.} \label{IIPcomp_L}
\end{figure}

As before, the effect of the filter is best shown by a comparison of the spectra before and after the filter, Fig. \ref{IISpcomp_L}.  The long, monochromatic tail in the power distribution on the left side of Fig. \ref{IIPcomp_L} constitutes the time-domain effect of the filtering procedure.

\begin{figure}[tb]
\includegraphics[width=1.0\textwidth]{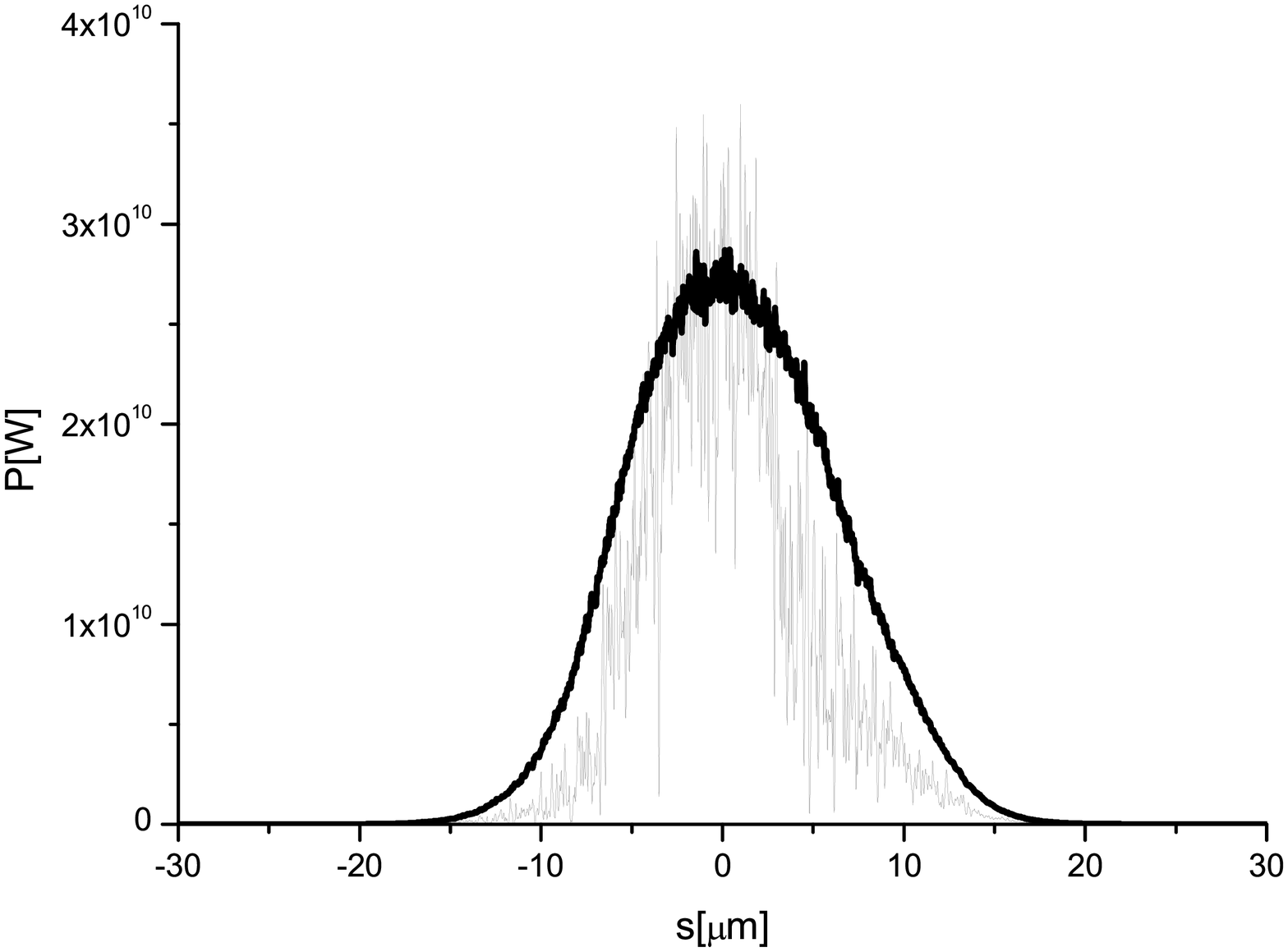}
\caption{Three-undulator configuration. Long pulse mode of operation. Average and typical single-shot output power (respectively, thick and thin solid lines).} \label{Pout_L}
\end{figure}

\begin{figure}[tb]
\includegraphics[width=1.0\textwidth]{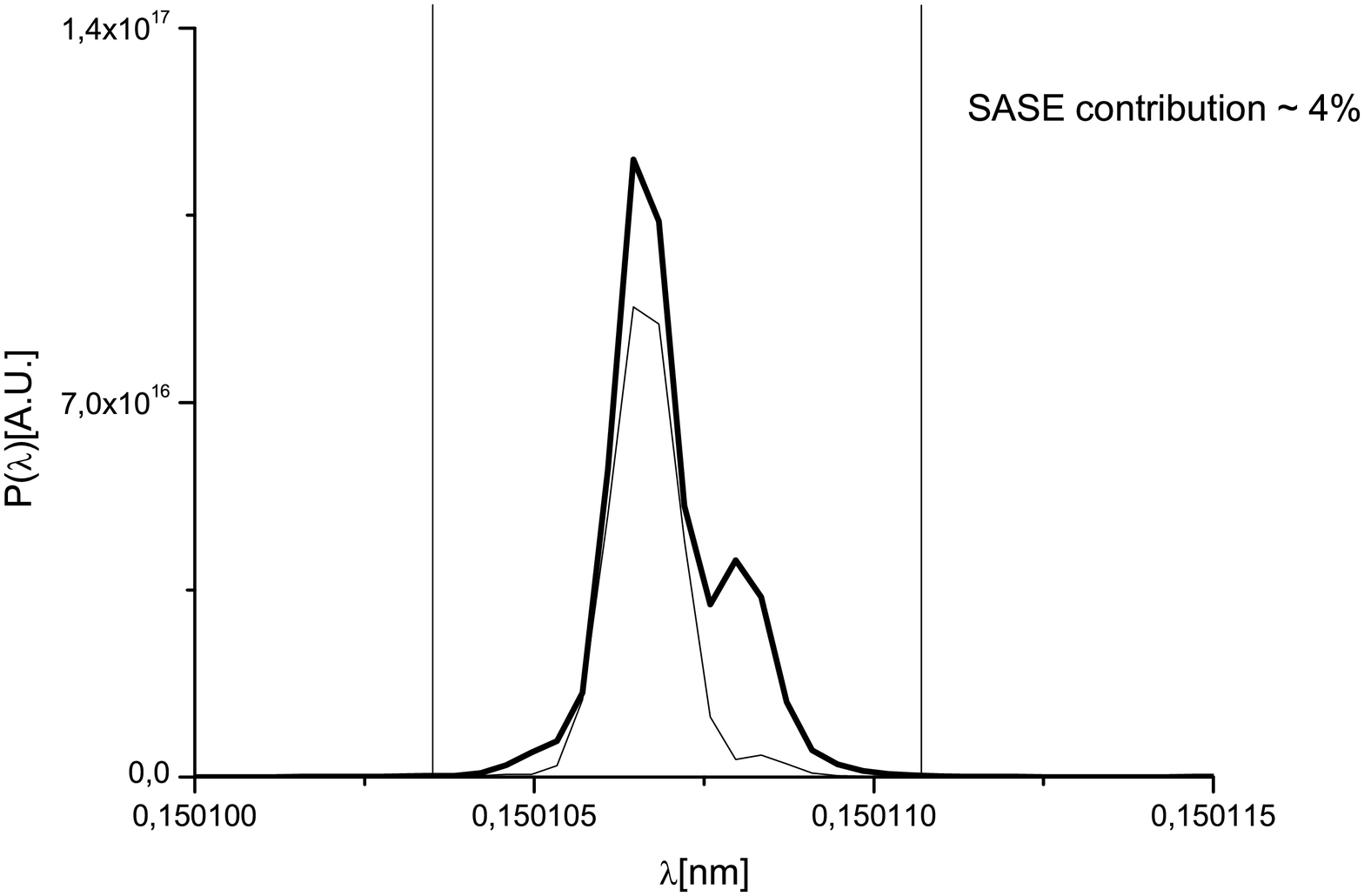}
\caption{Three-undulator configuration. Long pulse mode of operation. Average and typical single-shot output spectrum (respectively, thick and thin solid lines). An estimation of the SASE contribution can be done by evaluating the total power outside the spectral window shown with black straight lines and dividing it to the spectral power integrated over all the spectrum.} \label{Spout_L}
\end{figure}

\begin{figure}[tb]
\includegraphics[width=1.0\textwidth]{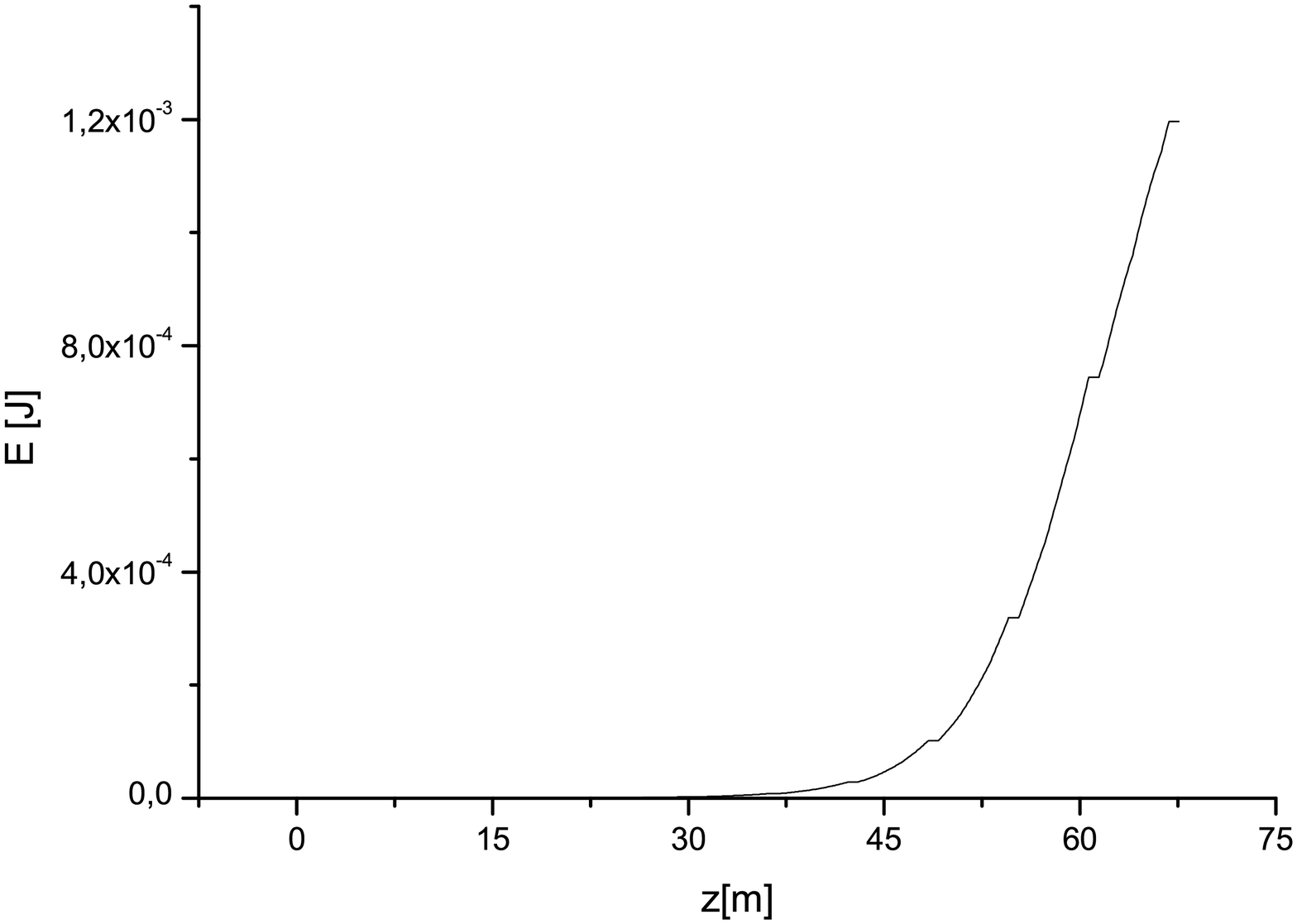}
\caption{Three-undulator configuration.  Long pulse mode of operation. Average energy of the radiation pulse as a function of the third undulator length.} \label{eneL}
\end{figure}

\begin{figure}[tb]
\includegraphics[width=1.0\textwidth]{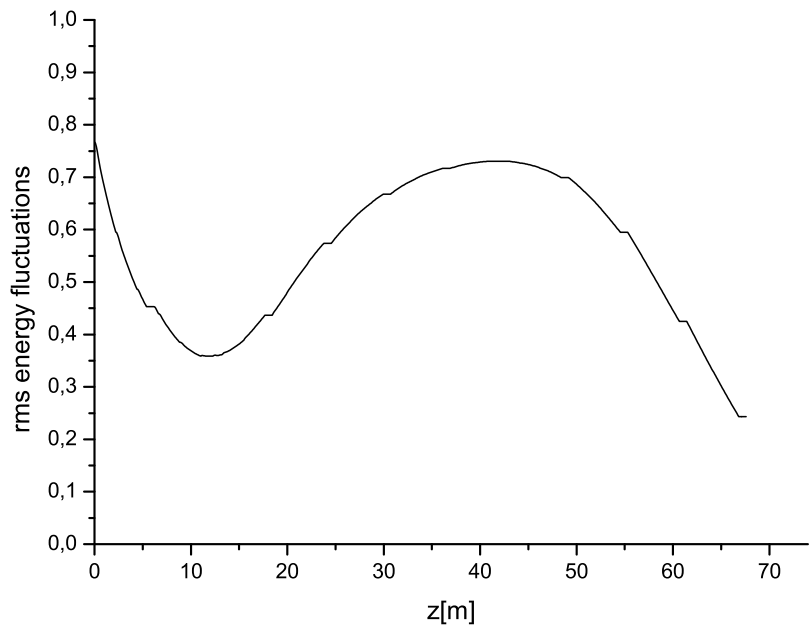}
\caption{Three-undulator configuration. Long pulse mode of operation. rms energy deviation from the average as a function of the third undulator length.} \label{rmsL}
\end{figure}

Following the second crystal, the radiation is used to seed once more the electron bunch. Radiation is collected at the exit of a third undulator. Fig. \ref{Pout_L} and Fig. \ref{Spout_L} respectively show the output power and spectrum for the three-undulator configuration, while in Fig. \ref{eneL} and Fig. \ref{rmsL} we present, respectively, the average energy of the pulse as a function of the undulator length, and the rms deviation from the average. The SASE contribution is strongly reduced to a few percent.


\clearpage

\section{Conclusions}

The fundamental problem of reducing the line width of SASE
X-ray FELs is solved by implementing a
self-seeding technique. In the present work, we described how to avoid the poor longitudinal coherence of hard X-ray SASE pulses. Quite surprisingly, monochromatization can be performed by an almost trivial setup composed of as few as two components: a weak chicane, and a single crystal. We have described techniques for
reducing the line width of the output X-ray beam down to $10^{-5}$, i.e. down to Fourier-Transform limit of the radiation pulse. We have shown how to achieve this monochromatization with small heat-loading of monochromators, which  is crucial for the European XFEL. Many interesting applications can be found for these self-seeding techniques. However, in order keep this paper within a reasonable size, we did not discuss, here, possible
applications of the proposed method, leading e.g. to further
improvement of the XFEL performance. The proposed  cascade self-seeding scheme based on the use of the wake monochromator is
extremely compact and takes almost no cost and time to be implemented. It can be straightforwardly installed in the baseline undulator system of the European XFEL and is safe, in the sense that
it guarantees the baseline mode of operation.

\section{Acknowledgements}

We are grateful to Massimo Altarelli, Reinhard Brinkmann,
Serguei Molodtsov and Edgar Weckert for their support and their interest during the compilation of this work.


\begin{thebibliography}{99}

\bibitem{tdr-2006} M. Altarelli, et al. (Eds.)
XFEL, The European X-ray Free-Electron Laser, Technical Design
Report, DESY 2006-097, Hamburg (2006).

\bibitem{LCLS1} J. Arthur et al. (Eds.) Linac Coherent Light Source
(LCLS). Conceptual Design Report, SLAC-R593, Stanford (2002) (See
also http://www-ssrl.slac.stanford.edu/lcls/cdr).

\bibitem{LCLS2} P. Emma, First lasing of the LCLS X-ray FEL at 1.5 Å, in
Proceedings of PAC09, Vancouver, to be published in
http://accelconf.web.cern.ch/AccelConf/ (2009).

\bibitem{DING} Y. Ding et al., Phys. Rev. Lett. 102, 254801
(2009).

\bibitem{SPRIN} T. Tanaka et al. (Eds.) Spring-8 Compact SASE Source Conceptual Design report, Kouto
(2005) (See also http://www-xfel.spring8.or.jp/SCSSCDR.pdf)

\bibitem{SELF} J. Feldhaus et al., Optics. Comm. 140, 341 (1997).


\bibitem{SXFE} E. Saldin, E. Schneidmiller,  Yu. Shvyd'ko and M.
Yurkov, NIM A 475 357 (2001).

\bibitem{OURL} G. Geloni, V. Kocharyan and E.~Saldin, "Scheme for generation of highly monochromatic X-rays from a baseline
XFEL  undulator", DESY 10-033 (2010).

\bibitem{HUAN} Y. Ding, Z. Huang and R. Ruth, Proceedings of the FLS 2010, ICFA Beam Dynamic Workshop, Menlo Park, to be published.

\bibitem{OURX} G. Geloni, V. Kocharyan and E.~Saldin, "A simple method for controlling the line width of SASE X-ray FELs",
DESY 10-053 (2010).

\bibitem{HUAYU} I. Ben-Zvi and L.H. Yu, Nucl. Instr. and Meth. A
393, 96 (1997).

\bibitem{SAL1} E. Saldin, E. Schneidmiller and M. Yurkov, Opt.
Commun. 212, 377 (2002).

\bibitem{SAL2} E. Saldin, E. Schneidmiller and M. Yurkov, Opt.
Commun., 239, 161 (2004).

\bibitem{OUR01} G. Geloni, V. Kocharyan and E.~Saldin, "Scheme for femtosecond-resolution pump-probe
experiments at XFELs with two-color ten GW-level X-ray pulses",
DESY 10-004 (2010).

\bibitem{OUR02} G. Geloni, V. Kocharyan and E.~Saldin, "The potential for extending the spectral range
accessible to the European XFEL down to 0.05 nm", DESY 10-005
(2010).

\bibitem{OUR03} G. Geloni, V. Kocharyan and E.~Saldin, "Scheme for simultaneous generation of three-color  ten GW-level X-ray
pulses from baseline XFEL undulator and multi-user distribution
system for XFEL laboratory", DESY 10-006 (2010).

\bibitem{GENE} S Reiche et al., Nucl. Instr. and Meth. A 429, 243 (1999).


\end{thebibliography}
\end{document}